\documentclass[11pt]{article}
\usepackage[dvips]{epsfig}
\usepackage{amsfonts,amsmath,amsbsy,xspace}
\makeatletter
\def\upA{\uparrow}
\def\dnA{\downarrow}

\def\B{{\cal B}}

\def\H{{\cal H}}

\def\L{{\cal L}}

\def\P{{\cal P}}
\def\O{{\cal O}}
\def\OO{\widehat{O}}

\def\a{{\cal A}}

\def\mg{\ell_{B}^{2}}
\def\dx{d^{2}x}
\def\dy{d^{2}y}

\def\dq{d^{2}q}

\def\CP{CP$^1$ }

\def\llangle{\langle\!\!\langle}
\def\rrangle{\rangle\!\!\rangle}

\def\Wed#1#2{\mg{#1\!\wedge\!#2\over2}}
\def\bpmatrix{\left(\begin{array}{c}}
\def\epmatrix{\end{array}\right)}
\def\F{{\mathfrak S}}
\def\FF{\widetilde{\F}}
\def\Laugh{\F_{\text{LN}}[\bbox{x}]}
\def\CBs{composite bosons\xspace}

\def\CB{composite-boson\xspace}
\def\bbox#1{\boldsymbol#1}
\def\LLL{lowest\ Landau\ level\xspace}
\@addtoreset{equation}{section}

\def\epsFbox[#1]#2#3#4{
\begin{figure}[!bt]
\begin{center}\small\epsfbox{#2}\label{#3}\caption{#4}\end{center}\end{figure}}
\def\FigOut#1{#1}
\def\FigFQHE1{
\epsFbox[!htb]{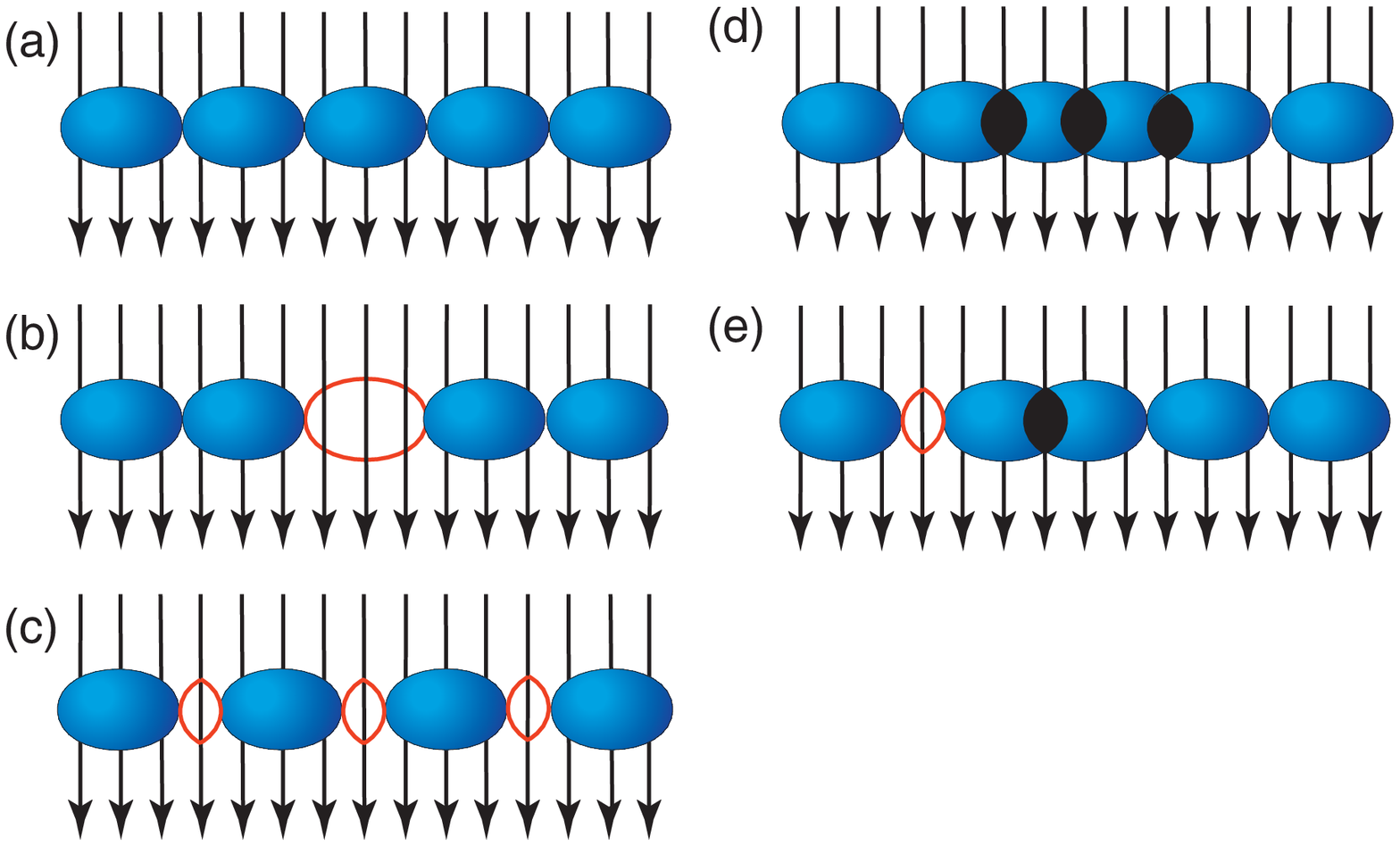}{FQHE1PS}{%
We illustrate the fractional QH state at $\nu =1/3$.  There are $3$ flux quanta 
per electron.
(a) The ground state is a closed packet of electrons pierced by 3 flux quanta.
(b) When one electron is removed one big hole is created.
(c) To lower the Coulomb energy there appear 3 small holes, which act as 
quasiholes.
(d) When one electron is added there appear 3 overlapping of electrons, which 
act as quasielectrons.
(e) Thirmal fluctuations create quasihole-quasielectron pairs.
}
}
\def\FigSkyrExci{
\epsFbox[!htb]{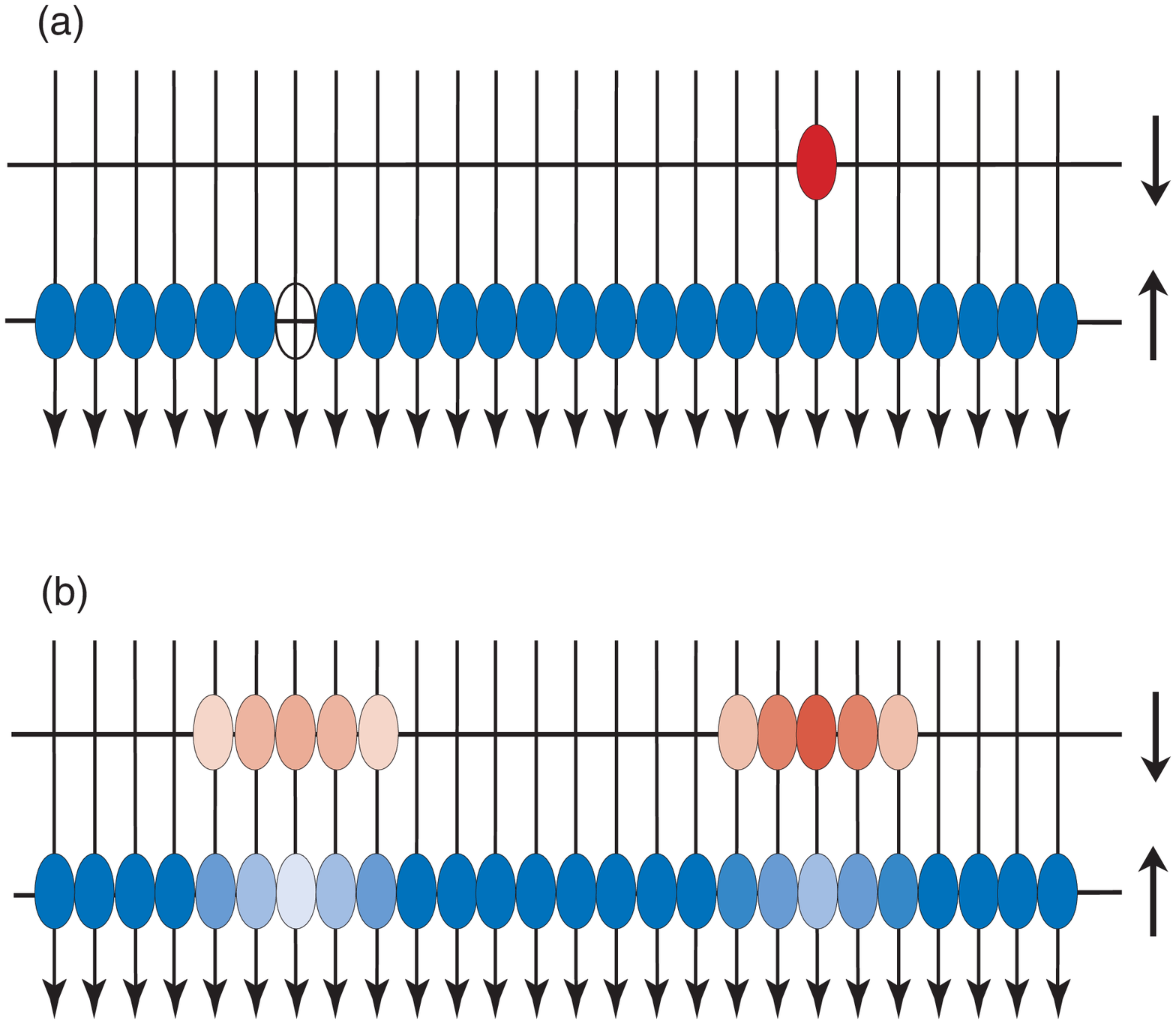}{SkyrExciPS}{%
We illustrate the QH ferromagnet at $\nu =1$.  There are one flux quantum per 
electron.  The ground state is a closed packet of electrons pierced by one 
flux quantum in the spin-up state.
(a) When one electron is removed one hole is created, while when one electron 
is added it is placed in the spin-down state.  Their activation energy is 
$\sim e^{2}/\varepsilon \ell _{B}$.
(b) To lower the Coulomb energy neighboring electrons make spin rotations  
The spin rotation modulates the electron density coherently and
spreads the charge to a wider domain of size $\kappa \ell _{B}$, 
as decreases the Coulomb energy by factor $1/\kappa $.  
Quasiparticles are coherent excitations called Skyrmions.
}
}
\def\FigVortNume{
\epsFbox[!htb]{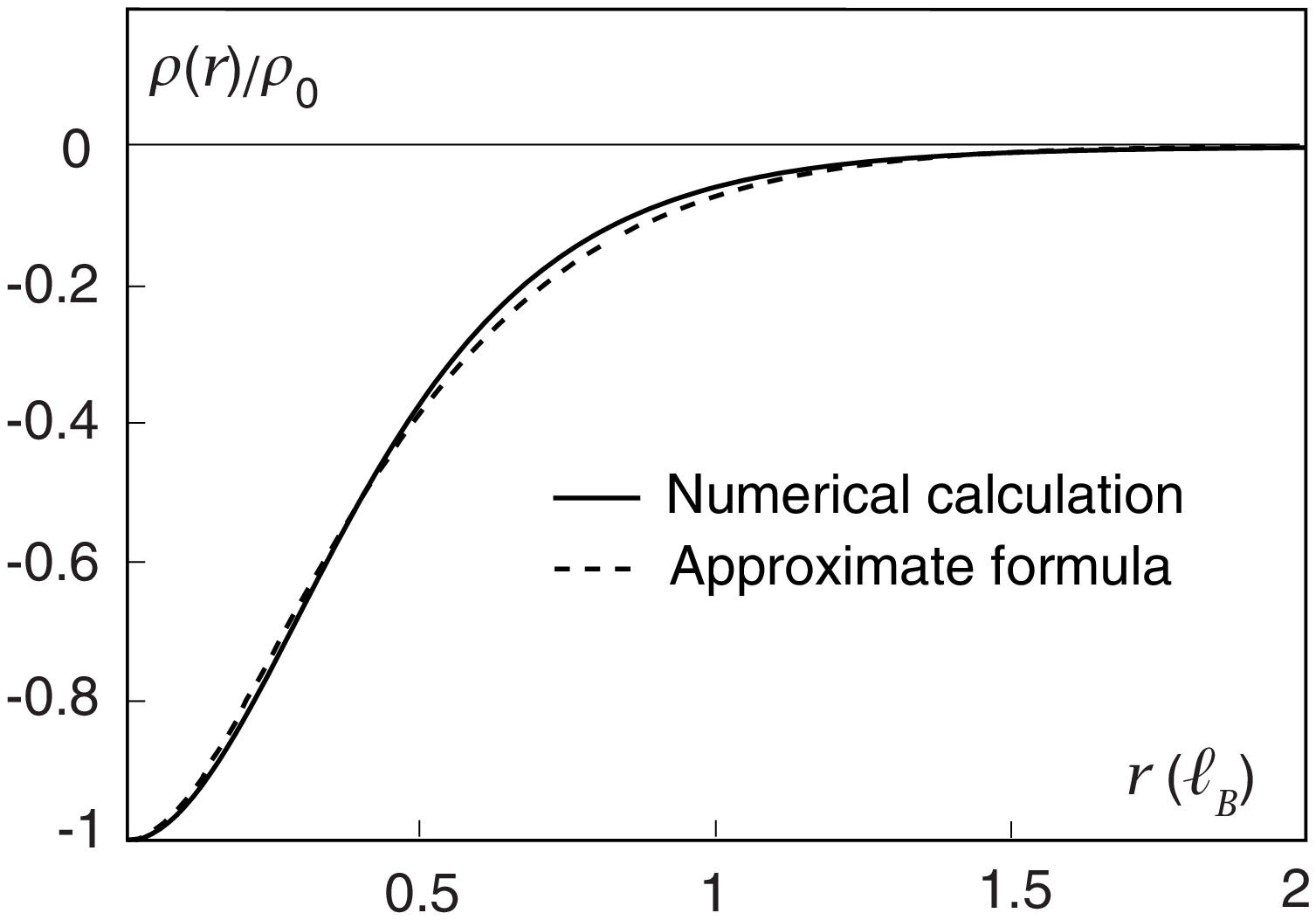}{VortNumePS}{
The density moduration around a vortex with $q=1$ is plotted.  The solid curve 
is obtained by solving the differential equation (\ref{ModifLiuvi})  
numerically.
The dashed curve is drawn by using the approximate formula (\ref{BetteAppro}).  
}
}
\def\FigSkyrEneU{
\epsFbox[!htb]{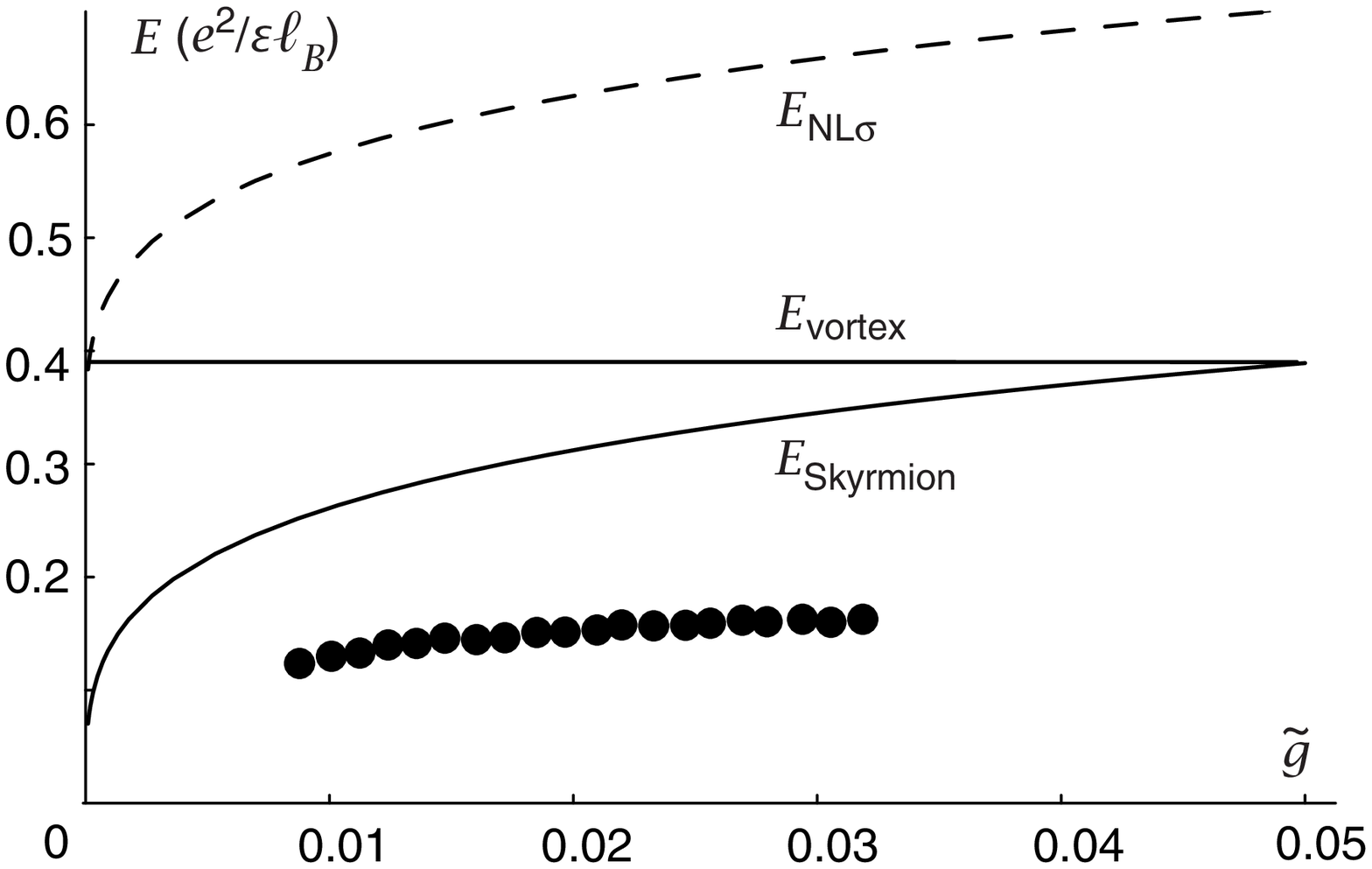}{SkyrEneUPS}{
The activation energy is represented by the thick curve for one Skyrmion 
($E_{\text{Skyrmion}}$) and by the thin horizontal line for one vortex 
($E_{\text{vortex}}$) in unit of $e^{2}/\varepsilon \ell _{B}$.  The dashed curve represents the 
Skyrmion activation energy based on the NL$\sigma $ model (\ref{TheirFormu}).  The filled 
circles are taken from the experimental data due to Schmeller et al. 
normalized for one quasiparticle excitation.  
}
}
\def\FigSkyrEneT{
\epsFbox[!htb]{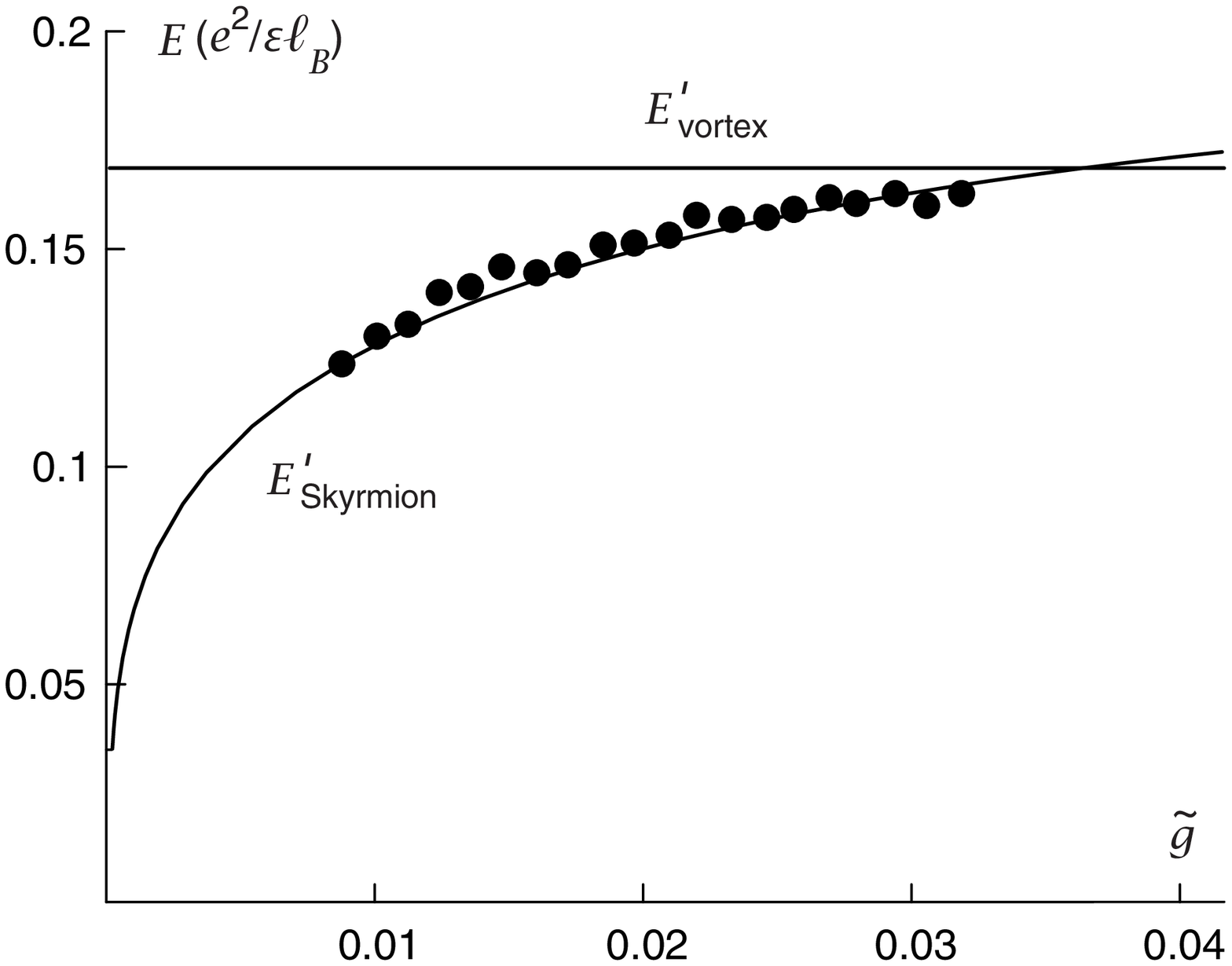}{SkyrEneTPS}{
The best fit for the observed Skyrmion excitation energy is obtained provided 
the Coulomb energy is decreased by 54\%, as plotted by the curve for 
$E'_{\text{Skyrmion}}$ and the horizonatal line for $E'_{\text{vortex}}$.
}
}
\makeatother
\begin{document}
\title{Field Theory of Quantum Hall Effects, \\
Composite Bosons, Vortices and Skyrmions}
\author{Z.F. Ezawa}
\date{Department of Physics, Tohoku University, Sendai 980-8578, Japan}
\maketitle
\begin{abstract}
\textwidth=140mm \renewcommand{\baselinestretch}{1.3}
A field theory of quantum Hall effects is constructed based on the \CB 
picture.  It is tightly related with the microscopic wave-function theory.  
The characteristic feature is that the field operator describes solely the 
physical degrees of freedom representing the deviation from the Laughlin 
state.  It presents a powerful tool to analyze excited states within the \LLL.  
It is shown that all excitations are nonlocal topological solitons in the 
spinless quantum Hall system.  On the other hand, in the presence of the spin 
degree of freedom it is shown that a quantum coherence develops spontaneously, 
where excitations include a Goldstone mode besides nonlocal topological 
solitons.  Solitons are vortices and Skyrmions carrying the U(1) and SU(2) 
topological charges, respectively.  Their classical configurations are derived 
from their microscopic wave functions.  The Skyrmion appears merely as a 
low-energy excitation within the \LLL and not as a solution of the effective 
nonlinear sigma model.  We use it as a consistency check of the Skyrmion 
theory that the Skyrmion is reduced to the vortex in the vanishing limit of 
the Skyrmion size.  We evaluate the activation energy of one Skyrmion and 
compare it with experimental data.\\
\end{abstract}

\section{Introduction}

The quantum Hall (QH) effect \cite{FQHEbook} is a remarkable macroscopic 
quantum phenomenon observed in the two-dimensional electron system at low 
temperature $T$ and in strong magnetic field $B$.  The Hall conductivity is 
quantized with extreme accuracy and develops a series of plateaux at magic 
values of the filling factor $\nu =2\pi \hbar \rho _{0}/eB$.  Here, $\rho _{0}$ is the average electron 
density.  The longitudinal resistivity is
\begin{equation}
\rho _{xx} \propto  \exp\Bigl(-{\Delta \over 2k_{B}T}\Bigr) ,
\label{LongiResis}
\end{equation}
with $k_{B}$ the Boltzman constant and $\Delta $ the activation energy of a 
quasihole-quasielectron pair, $\Delta =\Delta _{h}+\Delta _{e}$.

It is widely recognized that the QH effect comes from the realization 
of an incompressible ground state.  The \CB picture has proved to be quite 
useful to understand all essential aspects of QH effects 
\cite{GirvinMc,ZhangHaKi,ReadA,EHIf}.  Electrons may condense into an 
incompressible quantum Hall liquid as composite bosons.  The QH state is such 
a condensate of composite bosons, where quasiparticles are vortices 
\cite{LaughlinA}.  When the SU(2) symmetry is incorpolated, a quantum coherence 
may develop spontaneously \cite{EIcoher,EzaIQC,MG}, turning the QH system into a 
QH ferromagnet.  New excitations are Skyrmions \cite{SkyrmQH,SkyrmExper}.  They 
were initially introduced based on the \CB picture \cite{SkyrmQH} and later 
studied also in a microscopic Hartree-Fock approximation \cite{SkyrmHartFock}.  
However, the \CB theory is not satisfactory at all.  The naive formulation 
suffers from a serious problem we mention later.  Furthermore, the relation 
between the \CB theory and the microscopic wave-function theory is unclear.  
For instance, although the Skyrmion wave function is reduced to the vortex 
wave function in the vanishing limit of the Skyrmion size \cite{MG}, this property 
is lost for the Skyrmion and vortex classical fields in the effective 
nonlinear sigma (NL$\sigma $) model of Skyrmions \cite{SkyrmQH,MG}.  The aim of this 
paper is to present an improved scheme of \CBs free from these difficulties.

The idea of the improved \CB theory is summarized as follows.  For the 
sake of simplicity we consider the QH system neglecting the spin degree of 
freedom.  We use the complex coordinate normalized as $z=(x+iy)/2\ell _{B}$ with $\ell _{B}$ 
the magnetic length.  We are concerned about physics taking place within the 
\LLL.  Indeed, at sufficiently low temperature the relevant excitations are 
those confined within the \LLL.  At $\nu =1/m$ ($m$ odd), any state $|\F\rangle $ in the 
\LLL is described by a wave function,
\begin{equation}
\F[\bbox{x}] \equiv  \langle 0|\psi (\bbox{x}_{1})\cdots \psi (\bbox{x}_{N})|\F\rangle  = \omega [z]\Laugh ,
\label{WaveElect}
\end{equation}
with $\Laugh$ the Laughlin function \cite{LaughlinA},
\begin{equation}
\Laugh = \prod _{r<s}(z_{r}-z_{s})^{m}e^{-\sum _{r=1}^{N}|z_{r}|^{2}}, \label{LaughWave}
\end{equation}
where $\omega [z]\equiv \omega (z_{1},z_{2},\cdots ,z_{N})$ is an analytic function symmetric in all $N$ 
variables.  There is every reason \cite{FQHEbook} to believe that the ground state 
is the Laughlin state with $\F[\bbox{x}]=\Laugh$.  Since $\omega [z]$ is symmetric in $N$ 
variables, it must be a wave function of certain bosons, to which we refer as 
\textit{dressed composite bosons}.  Namely, we propose an improved 
bosonization by a mapping from the fermionic wave function $\F[\bbox{x}]$ to a 
bosonic wave function $\F_{\varphi }[\bbox{x}]$,
\begin{equation}
\F[\bbox{x}] \mapsto  \F_{\varphi }[\bbox{x}] \equiv  \omega [z].
\label{DressBosonMappi}
\end{equation}
Denoting its field operator by $\varphi (\bbox{x})$ we expect
\begin{equation}
\F_{\varphi }[\bbox{x}] \equiv  \langle 0|\varphi (\bbox{x}_{1})\cdots \varphi (\bbox{x}_{N})|\F\rangle  = \omega [z],
\label{CompoIntroC}
\end{equation}
which we establish in Section \ref{SecQHS}.  The ground state is extremely 
simple in terms of composite bosons, where $\F_{\varphi }[\bbox{x}]=1$.  The power of $z_{r}$ in 
$\omega [z]$ is the angular momentum carried by the $r$th composite boson.  Hence it 
is interpreted that the Laughlin state is the one where all composite bosons 
condense into the angular-momentum zero state.  Similarly, the vortex state 
with $\omega [z]=\prod _{r}z_{r}$ is the one where all \CBs condense into the angular-momentum 
one state.  

It is a characteristic feature of this theory that composite bosons 
represent the physical degree of freedom describing solely the deviation 
$\omega [z]$ from the ground state.  When the $N$-body wave function $\F_{\varphi }[\bbox{x}]$ is 
factorizable, $\omega [z]=\prod _{r}\omega (z_{r})$, it follows from (\ref{CompoIntroC}) that the 
one-point function is analytic,
\begin{equation}
\langle \varphi (\bbox{x})\rangle  = \omega (z) .
\label{OnePoint}
\end{equation}
This is a highly nontrivial requirement.  It connects directly the 
classical field $\langle \varphi (\bbox{x})\rangle $ of an excitation to its microscopic wave function 
$\omega (z)$.  In particular, it determines how the electron density modulates 
around the zeros of $\omega (z)$.  We demonstrate that all excitations are 
topological solitons (vortices) in the QH state when spins are neglected, and 
that there appears a Goldstone mode and topological solitons (vortices and 
Skyrmions) in the QH state when spins are taken into account.  Contrary to the 
common belief \cite{SkyrmQH}, the Skyrmion excitation appears merely as a 
low-energy excitation confined within the \LLL and not as a solution of the 
effective NL$\sigma $ model.  Because of this origin the Skyrmion is naturally 
reduced to the vortex in the vanishing limit of the Skyrmion size.  

The activation energy of the state $|\F\rangle $ is given by the matrix 
element, $\langle H\rangle =\langle \F|H|\F\rangle $, with $H$ the Hamiltonian.  The calculation is quite 
intriguing due to the condition that the excitation is confined within the 
\LLL.  We have two complementary methods.

One method is to use a semiclassical approximation, which is 
appropriate to obtain the activation energy of a topological soliton.  As we 
have stated, an investigation of the one-point function (\ref{OnePoint}) reveals 
how the densiy modulation (and also the spin modulation in the QH ferromagnet) 
is induced around the topological soliton when the excitation is confined 
within the \LLL.  The excitation energy is calculable with the knowledge of 
the density modulation (and the spin modulation).  

The other is an algebraic method based on the lowest-Landau-level (LLL) 
projection \cite{refLLL}.  The electron coordinate $\bbox{x}=(x,y)$ is decomposed into 
the guiding center (center-of-mass coordinate) $\bbox{X}=(X,Y)$ and the relative 
coordinate $\bbox{R}=(R_{x},R_{y})$.  Because the physics in the \LLL involves only the 
guiding center $\bbox{X}$, the symmetry of the two-dimensional space is subject to 
the magnetic translation group, which is generated by the magnetic translation 
$e^{i\bbox{q}\bbox{X}}$ and not by the Abelian translation $e^{i\bbox{q}\bbox{x}}$.  It has a crucial 
consequence \cite{EzaIQC,MG} that the spin density SU(2) operator and the electron 
density U(1) operator become noncommutative, because $X$ and $Y$ coordinates 
of the guiding center $\bbox{X}$ are noncommutative.  It implies that a spin rotation 
induces an electron density modulation and hence requires a Coulomb exchange 
energy, as is consistent with the semiclassical result.  Making a perturbative 
expansion of the spin texture around the ground state, we derive the NL$\sigma $ 
model from the matrix element $\langle \F|H|\F\rangle $ as the effective Hamiltonian 
describing perturbative spin fluctuations in QH ferromagnets \cite{EzaIQC,MG}.  
\FigOut{\FigFQHE1}

Before we present a detailed discussion of a field theory of \CBs, let 
us describe a physical picture of the fractional QH system.  We neglect the 
spin degree of freedom for simplicity.  All electrons are assumed to be in the 
\LLL.  Due to the Pauli exclusion principle only one electron can occupy one 
quantum-mechanical state.  The density of states in each energy level is 
$D_{n}=1/(2\pi \ell _{B}^{2})$, which is equal to the number of Dirac flux quanta passing 
through unit area, $D_{n}=B/\phi _{D}$, where the Dirac flux quantum is $\phi _{D}\equiv 2\pi \hbar /e$.  The 
filling factor of the energy level is thus defined by $\nu ={\rho _{0}/D_{n}}=2\pi \hbar \rho _{0}/eB$.  
It is characteristic that there are $m$ flux quanta per electron, $B/\rho _{0}=m\phi _{D}$, 
at $\nu =1/m$.  We are able to attach $m$ flux quanta to one electron by way of a 
phase transformation \cite{Wilczek}, composing a composite particle.  It is a 
composite fermion \cite{JainCF} for even $m$, while it is a composite boson for 
odd $m$.  Composite particles feel the effective magnetic field, which 
vanishes at $\nu =1/m$.  When $m$ is odd, composite bosons become free and 
undergo a bose condenstation at $\nu =1/m$.  We may view that the condensate is a 
closed packet of composite bosons pierced by $m$ flux quanta, as is 
illustrated in Fig.\ref{FQHE1PS}(a).  It is the nondegenerate ground state 
with homogeneous electron density.  When one electron is removed, one "big" 
hole would appear in the homogeneous electron density as in 
Fig.\ref{FQHE1PS}(b).  It represent a charge defect $e$ occupying three flux 
quanta.  It is energitically favorable that it is dissociated into $m$ "small" 
holes, each of which represents charge defect $e/m$ pierced by one flux 
quantum, as in Fig.\ref{FQHE1PS}(c).  The small hole is not smeared out since 
it is combined with a flux quantum.  Indeed, it is the vortex soliton 
(quasihole) carrying electric charge $e/m$, as we argue in the body.  When one 
electron is added, there appears $m$ overlapping of flux-electron composites 
as in Fig.\ref{FQHE1PS}(d).  Each of which represents charge excess $-e/m$ 
occupying one flux quantum.  It is the antivortex soliton (quasielectron) 
carrying electric charge $-e/m$.  Thirmal fluctuations generate 
quasihole-quasielectron pairs, as is illustrated in Fig.\ref{FQHE1PS}(e).

One quasielectron (quasihole) carries the electric charge $-e/m$ 
($+e/m$) and the magnetic flux $-\phi _{D}$ ($+\phi _{D}$) within the domain of size $\ell _{B}$.  
Hence, the creation energy $\Delta _{e}$ ($\Delta _{h}$) of a quasielectron (quasihole) is 
$\sim e^{2}/(m^{2}\varepsilon \ell _{B})$.  The system is said to be incompressible when there is a gap in 
the chemical potential as a function of the electron density.  The fractional 
QH system is incompressible because the gap of the chemical potential is 
$m(\Delta _{e}+\Delta _{h})$ at $\nu =1/m$.
\FigOut{\FigSkyrExci}

We next gives a physical picture of the QH ferromagnet at $\nu =1$.  When 
the Zeeman effect is small, each Landau level contains two almost degenerate 
levels with spin-up and spin-down states.  All electrons are in the spin-up 
state.  There is one flux quantum per electron, and the ground state is a 
closed packet of composite bosons pierced by one flux quantum.  It is the 
nondegenerate ground state with homogeneous electron density.  (If the Zeeman 
effect is absent, it is one of degenerate ground states.)  When one electron 
is removed, one hole appears as in Fig.\ref{SkyrExciPS}(a).  It is a vortex 
just as in the spinless QH state in Fig.\ref{FQHE1PS}.  The Coulomb energy is 
$\sim e^{2}/\varepsilon \ell _{B}$.  When one electron is added, it is placed in the spin-down state as 
in Fig.\ref{SkyrExciPS}(a).  It is a localized lump of a quantized electric 
charge, and its Coulomb energy is also $\sim e^{2}/\varepsilon \ell _{B}$.  This is the picture of the 
integer QH state when the Zeeman energy is very large.  When the Zeeman energy 
is negligible, however, it is possible to lower the Coulomb energy by rotating 
the spins of neighboring electrons, as in Fig.\ref{SkyrExciPS}(b).  When the 
spin rotates, the electron density is also modulated coherently because the 
SU(2) and U(1) operators are noncommutative due to the 
magnetic-translation-group effect.  As a result the electric charge is smeared 
into a wider domain.  This is made possible by developing quantum coherence 
spontaneously.  Resulting coherent excitations are Skyrmions.  When the size 
of the Skyrmion is $\kappa \ell _{B}$, the Coulomb energy is deduced by factor $1/\kappa $.  The 
size is determined by the competition between the Coulomb energy and the 
Zeeman energy.  It is clear in this picture that the Skyrmion 
[Fig.\ref{SkyrExciPS}(b)] is reduced to the vortex [Fig.\ref{SkyrExciPS}(a)] 
when the Zeeman energy is sufficiently large.

This paper is composed as follows.  
In Section \ref{SecBoson} \CB fields are defined.  First, as in the standard 
\CB theory \cite{ZhangHaKi,EHIf}, we attach an odd number of flux quanta to an 
electron by way of a singular phase transformation \cite{Wilczek}.  We call the 
resulting electron-flux composite the \textit{bare composite boson}.  In order 
to soften the singularity brought in, we dress it with a cloud of an effective 
magnetic field that bare \CBs feel.  The resulting object turns out to be the 
\textit{dressed composite boson}.  
In Section \ref{SecQHS} the relation is established between the electron wave 
function (\ref{WaveElect}) and the \CB wave function (\ref{CompoIntroC}).  We also 
verify that the ground state is given by the Laughlin state within the 
semiclassical approximation.  
In Section \ref{SecTE} we make a semiclassical analysis of vortex excitations 
on the basis of the formula (\ref{OnePoint}).

In Section \ref{SecQHFerro} we define \CB fields with the spin degree 
of freedom.  It is shown that quantum coherence develops spontaneously when 
the Zeeman effect is small.  We review on the Goldstone mode characterizing 
the QH ferromagnet.  
In Section \ref{SecSpinTE} we discuss vortex and Skyrmion excitations in the 
QH ferromagnet.  The Skyrmion classical configuration is also derived directly 
from its microscopic wave function based on the generalized formula of 
(\ref{OnePoint}).  We evaluate the activation energy of one Skyrmion and compare it 
with experimental data.  Throughout the paper we use the natural unit $\hbar =c=1$.
\section{Bosonization}\label{SecBoson}

The field-theoretical Hamiltonian for spinless planar electrons in 
external magnetic field $(0,0,-B)$ is given by
\begin{subequations}
\begin{align}
H&= {1\over 2M}\int \dx \psi ^{\dagger }(\bbox{x})(P_{x}^{2}+P_{y}^{2})\psi (\bbox{x}) + H_{C}    \label{Hamil:A}\\
&= {1\over 2M}\int \dx \psi ^{\dagger }(\bbox{x})(P_{x}-iP_{y})(P_{x}+iP_{y})\psi (\bbox{x}) + {N\over 2}\omega _{c} + H_{C} , \label{Hamil:B}
\end{align}
\end{subequations}
where $\psi (\bbox{x})$ is the electron field; $P_{j}=-i\partial _{j}+eA^{\text{ext}}_{j}$ is the covariant 
momentum with $A^{\text{ext}}_{j}={1\over 2}\varepsilon _{jk}x_{k}B$; $N$ is the electron number and $\omega _{c}$ 
is the cyclotron frequency.  The Coulomb interaction term is
\begin{equation}
H_{C} = {e^{2}\over 2\varepsilon }\int \dx\dy {\varrho  (\bbox{x})\varrho  (\bbox{y})\over |\bbox{x}-\bbox{y}|} , 
\label{CouloEnergPre}
\end{equation}
where $\varrho  (\bbox{x})\equiv \rho (\bbox{x})-\rho _{0}$ stands for the deviation of the electron density 
$\rho (\bbox{x})\equiv \psi ^{\dagger }(\bbox{x})\psi (\bbox{x})$ from its average value $\rho _{0}$.

The state $|\F\rangle $ in the \LLL obeys
\begin{equation}
(P_{x}+iP_{y})\psi (\bbox{x})|\F\rangle =-{i\over \ell _{B}}\Bigl(z + {\partial \over \partial z^{*}}\Bigr)\psi (\bbox{x})|\F\rangle  = 0 ,
\label{LLLcondiElect}
\end{equation}
upon which the kinetic Hamiltonian is trivial.  We call it the LLL condition.  
The generic solution of this equation yields the $N$-body wave function 
(\ref{WaveElect}) for the state $|\F\rangle $.  We are concerned about the state $|\F\rangle $ in 
the \LLL.

We start with a review of the standard bosonization scheme pioneered by 
Girvin and MacDonald \cite{GirvinMc}.  Its Landau-Ginzburg theory was proposed by 
Zhang et al. \cite{ZhangHaKi,MG} and its microscopic field theory was developped 
by Ezawa et al. \cite{EHIf,EIcoher}.  We define the field $\phi (\bbox{x})$ by an operator 
phase transformation,
\begin{equation}
\phi (\bbox{x}) = e^{-i\Theta (\bbox{x})}\psi (\bbox{x}) ,
\label{NaiveField}
\end{equation}
where $\Theta (\bbox{x})$ is the phase field,
\begin{equation}
\Theta (\bbox{x}) = m\int \dy \theta (\bbox{x}-\bbox{y})\rho (\bbox{y}) ,
\label{PhaseTheta}
\end{equation}
with the azimuthal angle $\theta (\bbox{x}-\bbox{y})$.  When $m$ is an odd integer, we can prove 
that $\phi (\bbox{x})$ is a bosonic operator.  Let us call the underlying boson the bare 
composite boson.  It is a hardcore boson satisfying the exclusion principle, 
$\phi (\bbox{x})^{2}=0$.  The LLL condition (\ref{LLLcondiElect}) reads
\begin{equation}
(\check{P}_{x}+i\check{P_{y}})\phi (\bbox{x})|\F\rangle =0 ,
\label{LLLcondiBare}
\end{equation}
where $\check{P_{j}}\equiv P_{j}+\partial _{j}\Theta (\bbox{x})$ is the covariant momentum for the bare \CB 
field.  By solving this condition \cite{EHIf,GirvinMcFi}, the $N$-body wave 
function is found to be $\F_{\phi }[\bbox{x}] \equiv  \omega [z]|\Laugh|$.  Namely, the standard 
bosonization is a mapping from the fermionic wave function $\F[\bbox{x}]$ to a 
bosonic wave function $\F_{\phi }[\bbox{x}]$,
\begin{equation}
\F[\bbox{x}] \mapsto  \F_{\phi }[\bbox{x}] \equiv  \omega [z]|\Laugh|. \label{NaiveBoson}
\end{equation}
The mapping is to attach $m$ Dirac flux quanta $m\phi _{D}$ to each electron by way 
of the phase transformation (\ref{NaiveField}), where $\phi _{D}=2\pi /e$.  Thus, the ground 
state of bare composite bosons is described by the modulus of the Laughlin 
function, $|\Laugh|$.  

Though the essential physics of the QH effect is revealed by this naive 
\CB theory \cite{GirvinMc,ZhangHaKi,EHIf,EzaIQC,MG}, it is a really complicated 
theory as the complicated ground-state wave function indicates.  Furthermore, 
the phase transformation (\ref{NaiveField}) brings in singularities.  In the basis 
where the electron position is diagonalized, $\rho (\bbox{x})=\sum _{r}\delta (\bbox{x}-\bbox{x}_{r})$, we obtain
\begin{equation}
e^{i\Theta (\bbox{x})} = \prod _{r}e^{i\theta (\bbox{x}-\bbox{x}_{r})},
\label{SinguPhase}
\end{equation}
which is singular at $\bbox{x}=\bbox{x}_{r}$.  These are the reasons why its semiclassical 
analysis suffers from problems at least in the lowest order approximation.  
Indeed, it yields the semiclassical ground-state wave function very different 
from the Laughlin wave function; See (\ref{NaiveGrounWave}) in a later section.  

We wish to introduce another \CB field $\varphi (\bbox{x})$ which makes the LLL 
condition as simple as possible.  We set
\begin{equation}
\varphi (\bbox{x}) = e^{-\a(\bbox{x})-i\Theta (\bbox{x})}\psi (\bbox{x}) ,
\label{DressField}
\end{equation}
together with an operator $\a(\bbox{x})$ to be determined later.  By substituting 
this into the LLL condition (\ref{LLLcondiElect}), provided $\a(\bbox{x})$ satisfies
\begin{equation}
\partial _{j}\a(\bbox{x}) = \varepsilon _{jk}\bigl\{\partial _{k}\Theta (\bbox{x}) + eA_{k}(\bbox{x})\bigr\} = \varepsilon _{jk}\partial _{k}\Theta (\bbox{x}) - {1\over 2\ell _{B}^{2}}x_{j} ,
\label{EqForA}
\end{equation}
it is reduced to a simple formula,
\begin{equation}
(\P_{x}+i\P_{y})\varphi (\bbox{x})|\F\rangle =-{i\over \ell _{B}}{\partial \over \partial z^{*}}\varphi (\bbox{x})|\F\rangle  = 0 ,
\label{LLLcondiDress}
\end{equation}
where $\P_{j}=P_{j}+\partial _{j}\Theta (\bbox{x})+\partial _{j}\a(\bbox{x})$ is the covariant momentum for the new \CB field 
$\varphi (\bbox{x})$.  Eq.(\ref{EqForA}) is easily solved as
\begin{equation}
\a(\bbox{x})= m\int \dy \ln\Bigl({|\bbox{x}-\bbox{y}|\over 2\ell _{B}}\Bigr) \rho (\bbox{y}) -{|z|^{2}} .
\label{DressA}
\end{equation}
In the basis with the electron position diagonalized, we find
\begin{equation}
e^{\a(\bbox{x})} \propto  \prod _{r}|\bbox{x}-\bbox{x}_{r}|^{m} e^{-|z|^2} ,
\end{equation}
which removes the singularities at $\bbox{x}=\bbox{x}_{r}$ in (\ref{SinguPhase}).  

The $N$-body wave function $\F_{\varphi }[\bbox{x}]$ of dressed composite bosons is 
obtained by solving the LLL condition (\ref{LLLcondiDress}), and it is given simply 
by an analytic function $\omega [z]$ as in (\ref{CompoIntroC}).  We shall verify in the 
next section that one analytic function $\omega [z]$ characterizes one state $|\F\rangle $ 
as in (\ref{WaveElect}) in terms of electrons, or as in (\ref{NaiveBoson}) in terms of 
bare composite bosons, or as in (\ref{CompoIntroC}) in terms of dressed composite 
bosons.  A type of the field operator (\ref{DressField}) was first considered by 
Read \cite{ReadA} in constructing a Landau-Ginzburg theory different from the one 
due to Zhang et al. \cite{ZhangHaKi}, and revived recently by Rajaraman et al. 
\cite{RajaramanCB}.  

The effective magnetic potential for the bare composite boson is 
$A^{\text{ext}}_{k}+(1/e)\partial _{k}\Theta $, which is rewritten as $(1/e)\varepsilon _{kj}\partial _{j}\a(\bbox{x})$.  Bare 
composite bosons feel the effective magnetic field $\B_{\text{eff}}(\bbox{x})$,
\begin{equation}
\B_{\text{eff}}(\bbox{x}) = e^{-1}\bbox{\nabla }^{2}\a(\bbox{x}) = m\phi _{D}\rho (\bbox{x}) - B.
\label{EffecMagne}
\end{equation}
The effective field vanishes, $\langle \B_{\text{eff}}\rangle =0$, on the homogeneous state 
$\langle \rho (\bbox{x})\rangle =\rho _{0}$ if $m\phi _{D}\rho _{0}=B$.  It occurs at the filling factor $\nu  \equiv  \rho _{0}\phi _{D}/B = 1/m$.  
At $\nu =1/m$ we rewrite the effective magnetic field as
\begin{equation}
e\B_{\text{eff}}(\bbox{x}) = \bbox{\nabla }^{2}\a(\bbox{x}) = 2\pi m \varrho  (\bbox{x}) ,
\label{StepA}
\end{equation}
with $\varrho  (\bbox{y})\equiv \rho (\bbox{y})-\rho _{0}$.  It is solved as
\begin{equation}
\a(\bbox{x}) = m\int \dy \ln\Bigl({|\bbox{x}-\bbox{y}|\over 2\ell _{B}}\Bigr) \varrho  (\bbox{y}) .
\label{SpinB}
\end{equation}
This formula is equivalent to (\ref{DressA}) at $\nu =1/m$.  Since the field $\varphi (\bbox{x})$ is 
constructed by dressing a cloud of the effective magnetic field, we have 
termed it the dressed \CB field.
\section{Quantum Hall States}\label{SecQHS}

We derive the Hamiltonian in terms of dressed composite bosons by 
substituting (\ref{DressField}) into (\ref{Hamil:B}),
\begin{equation}
H = {1\over 2M}\int d^{2}x \varphi ^{\ddag }(\bbox{x})(\P_{x}-i\P_{y})(\P_{x}+i\P_{y})\varphi (\bbox{x}) + {N\over 2}\omega _{c} + H_{C} ,
\label{HamilCB}
\end{equation}
where we have defined
\begin{equation}
\varphi ^{\ddag }(\bbox{x}) \equiv  \varphi ^{\dagger }(\bbox{x})e^{2\a(\bbox{x})} ,
\label{DressOperaB}
\end{equation}
with which $\rho (\bbox{x})=\psi ^{\dagger }(\bbox{x})\psi (\bbox{x})=\varphi ^{\ddag }(\bbox{x})\varphi (\bbox{x})$.  The covariant momentum 
$\P_{j}=P_{j}+\partial _{j}\Theta -\partial _{j}\a$ reads,
\begin{equation}
\P_{j} =  -i\partial _{j} + e(\delta _{jk} - i\varepsilon _{jk})\a_{k},\quad  \a_{k}(\bbox{x})=-{1\over e}\varepsilon _{kj}\partial _{j}\a(\bbox{x}).
\label{CovarMomenR}
\end{equation}
The Lagrangian density is
\begin{equation}
\L = \psi ^{\dagger }(i\partial _{t}-eA^{\text{ext}}_{t})\psi  - \H = \varphi ^{\ddag }(i\partial _{t}-eA^{\text{ext}}_{t}-\partial _{t}\Theta -\partial _{t}\a)\varphi  - \H ,
\end{equation}
where $\H$ is the Hamiltonian density and 
$A^{\text{ext}}_{\mu }=(A^{\text{ext}}_{t},A^{\text{ext}}_{k})$ is the potential of the external 
electromagnetic field.  The canonical conjugate of $\varphi (\bbox{x})$ is not $i\varphi ^{\dagger }(\bbox{x})$ but 
$i\varphi ^{\ddag }(\bbox{x})$.  Hence, the equal-time canonical commutation relations are
\begin{equation}
[\varphi (\bbox{x}), \varphi ^{\ddag }(\bbox{y})] = \delta (\bbox{x}-\bbox{y}),\qquad [\varphi (\bbox{x}), \varphi (\bbox{y})] = [\varphi ^{\ddag }(\bbox{x}), \varphi ^{\ddag }(\bbox{y})] = 0.
\label{CCRbosonD}
\end{equation}
They are also derived \cite{RajaramanCB} by an explicit calculation from those of 
the electron fields $\psi (\bbox{x})$ and $\psi ^{\dagger }(\bbox{x})$ based on the definition (\ref{DressField}). 

The composite-boson wave function is defined by
\begin{equation}
\F_{\varphi }[\bbox{x}]=\langle 0|\varphi (\bbox{x}_{1})\varphi (\bbox{x}_{2}) \cdots  \varphi (\bbox{x}_{N})|\F\rangle  .
\label{CompoStepADress}
\end{equation}
The LLL condition (\ref{LLLcondiDress}) implies that the wave function $\F_{\varphi }[\bbox{x}]$ is 
an analytic function, $\F_{\varphi }[\bbox{x}]=\omega [z]$.  With the use of the formula (\ref{DressA}) it 
is an easy exercise to derive the following relation \cite{RajaramanCB},
\begin{equation}
\varphi ^{\ddag }(\bbox{x}_{1})\varphi ^{\ddag }(\bbox{x}_{2})\cdots \varphi ^{\ddag }(\bbox{x}_{N})|0\rangle  = \Laugh \psi ^{\dagger }(\bbox{x}_{1})\psi ^{\dagger }(\bbox{x}_{2})\cdots \psi ^{\dagger }(\bbox{x}_{N})|0\rangle  ,
\label{DressB}
\end{equation}
where $\Laugh$ is the Laughlin function (\ref{LaughWave}).

Because of the commutation relations (\ref{CCRbosonD}) the state $|\F\rangle $ 
associated with the composite-boson wave function (\ref{CompoStepADress}) is given 
by
\begin{align}
|\F\rangle &= \int [d\bbox{x}] \F_{\varphi }[\bbox{x}] \varphi ^{\ddag }(\bbox{x}_{1})\varphi ^{\ddag }(\bbox{x}_{2}) \cdots  \varphi ^{\ddag }(\bbox{x}_{N})|0\rangle   \label{NbodyRb}\\
&= \int [d\bbox{x}] \omega [z]\Laugh \psi ^{\dagger }(\bbox{x}_{1})\psi ^{\dagger }(\bbox{x}_{2}) \cdots  \psi ^{\dagger }(\bbox{x}_{N})|0\rangle  , \label{NbodyRc}
\end{align}
where use was made of (\ref{DressB}), and $[d\bbox{x}]=\dx_{1}\dx_{2}\cdots \dx_{N}$.  It follows from 
(\ref{NbodyRc}) that the electron wave function is $\F[\bbox{x}]=\omega [z]\Laugh$, as verifies 
the basic formulas (\ref{WaveElect}) $\sim $ (\ref{CompoIntroC}).

One might question the hermiticity of the theory \cite{RajaramanCB} since 
the covariant momentum (\ref{CovarMomenR}) has an unusual expression.  It is 
related to the fact that the canonical conjugate of $\varphi (\bbox{x})$ is 
$i\varphi ^{\ddag }(\bbox{x})\equiv i\varphi ^{\dagger }(\bbox{x})e^{2\a(\bbox{x})}$.  It implies that the hermiticity is defined together 
with the measure $e^{2\a(\bbox{x})}$.  Such a measure has arisen since the 
transformation (\ref{DressField}) is not unitary.  It is trivial to check 
explicitly that the covariant momentum $\P_{j}$ is hermitian together with this 
measure.  It is also instructive to rewrite the Hamiltonian (\ref{HamilCB}) as
\begin{equation}
H = {\omega _{c}\over 2}\int \dx \Bigl({\partial \over \partial z^{*}}\varphi (\bbox{x})\Bigr)^{\dagger }e^{2\a(\bbox{x})}{\partial \over \partial z^{*}}\varphi (\bbox{x}) + {N\over 2}\omega _{c} + H_{C} ,
\label{HamilMono}
\end{equation}
which is manifestly hermitian.  

The semiclassical ground state is the one that minimizes the total 
energy $\langle H\rangle $.  The Coulomb energy (\ref{CouloEnergPre}) is minimized by the state 
where $\langle \varrho  (\bbox{x})\rangle =0$.  It is realized when the electron density is homogeneous, 
\begin{equation}
\langle \rho (\bbox{x})\rangle =\langle \varphi ^{\ddag }(\bbox{x})\varphi (\bbox{x})\rangle =e^{2\langle \a(\bbox{x})\rangle }\langle \varphi ^{\dagger }(\bbox{x})\varphi (\bbox{x})\rangle  =\rho _{0} .
\label{DressStepRho}
\end{equation}
In the semiclassical approximation we obtain
\begin{equation}
\prod _{r=1}^{N}\langle \rho (\bbox{x}_{r})\rangle  = \prod _{r=1}^{N}e^{2\langle \a(\bbox{x}_r)\rangle }\langle \F|\varphi ^{\dagger }(\bbox{x}_{N})\cdots \varphi ^{\dagger }(\bbox{x}_{2})\varphi ^{\dagger }(\bbox{x}_{1})\varphi (\bbox{x}_{1})\varphi (\bbox{x}_{2})\cdots \varphi (\bbox{x}_{N})|\F\rangle .
\end{equation}
We insert a complete set $\sum |n\rangle \langle n|=1$ between two operators $\varphi ^{\dagger }(\bbox{x}_{1})$ and 
$\varphi (\bbox{x}_{1})$.  When the state $|\F\rangle $ contains $N$ electrons, only the vacuum term 
$|0\rangle \langle 0|$ survives in the complete set because $\varphi (\bbox{x}_{r})$ decreases the electron 
number by one.  Hence, $N$-body wave function (\ref{CompoStepADress}) is given by
\begin{equation}
\F_{\varphi }[\bbox{x}] = \rho _{0}^{N/2} \prod _{r=1}^{N}e^{-\langle \a(\bbox{x}_r)\rangle } ,
\end{equation}
up to an irrelevant phase factor.  On the other hand, to suppress the kinetic 
energy we impose the LLL condition (\ref{LLLcondiDress}), as requires $\F_{\varphi }[\bbox{x}]$ to 
be analytic.  Consequently, it follows that $\langle \a(\bbox{x})\rangle =\text{constant}$ or 
$\B_{\text{eff}}=0$, which is possible only at $\nu =1/m$ from (\ref{EffecMagne}).  
Namely, the ground state is realized only at the magic filling factor $\nu =1/m$.  
At $\nu =1/m$ the ground-state wave function is given by $\F_{\varphi }[\bbox{x}]=\text{constant}$ 
in terms of \CBs, and therefore by the Laughlin wave function (\ref{LaughWave}) in 
terms of electrons.  In this way the Laughlin state is proved to be the ground 
state in the improved \CB theory.  The ground state is illustrated in 
Fig.\ref{FQHE1PS}(a).

For the sake of completeness, we briefly recall the results \cite{EHIf} of 
the corresponding analysis with use of bare composite bosons.  We can derive 
the equations similar to (\ref{NbodyRb}) and (\ref{NbodyRc}), which verifies the mapping 
(\ref{NaiveBoson}).  The semiclassical ground state is similarly given by 
$\F_{\phi }(\bbox{x})=\text{constant}$ in terms of bare composite bosons.  The wave function 
in terms of electrons is singular,
\begin{equation}
\F[\bbox{x}]=e^{im\sum _{r<s}\theta (z_{r}-z_{s})} . \label{NaiveGrounWave}
\end{equation}
The state does not belong to the \LLL.  Although the singular short-distance 
behavior in (\ref{NaiveGrounWave}) is remedied in a higher order perturbation 
theory \cite{EHIf,EIcoher}, it makes the naive composite-boson theory less 
attractive.  
\section{Semiclassical Analysis}\label{SecTE}

We analyze excitations on the QH state.  A priori two types of 
excitations are possible, that is, perturbative and nonperturbative ones in 
terms of the density fluctuation $\varrho  (\bbox{x})$ and its conjugate phase $\chi (\bbox{x})$.  A 
perturbative analysis has already been carried out based on the bare \CB 
theory \cite{EHIf}.  The result is that there exist no perturbative fluctuations 
confined within the \LLL.  This conclusion remains to be true in the improved 
theory.  To show  it, we parametrize the bare field as $\phi (\bbox{x}) = e^{i\chi (\bbox{x})} 
\sqrt {\rho _{0}+\varrho  (\bbox{x})}$ in terms of the density deviation $\varrho  (\bbox{x})$ and its canonical phase 
$\chi (\bbox{x})$.  The dressed field $\varphi (\bbox{x})$ is a nonlocal operator due to the factor 
$e^{-\a(\bbox{x})}$ with (\ref{SpinB}),
\begin{equation}
\varphi (\bbox{x}) = e^{-\a(\bbox{x})}e^{i\chi (\bbox{x})} \sqrt {\rho _{0}+\varrho  (\bbox{x})} .
\label{BareDress}
\end{equation}
Substituting (\ref{BareDress}) into the Hamiltonian (\ref{HamilCB}) and expanding 
various quantities in term of $\varrho  (\bbox{x})$ and $\chi (\bbox{x})$, the perturbative Hamiltonian 
is found to be identical between the bare and dressed \CB theory, as should be 
the case.  We conclude that all excitations in the \LLL are nonperturbative 
objects.

The improved theory confirms this assertion by showing explicitly how 
they are created on the ground state.  Indeed, any excited state is 
represented as in (\ref{NbodyRb}), which is a nonlocal object because the creation 
operator $\varphi ^{\ddag }(\bbox{x})$ is a nonlocal operator as in (\ref{BareDress}).
\subsection{Vortices}

We first examine excited states when the wave function 
(\ref{CompoStepADress}) is factorizable, $\F_{\varphi }[\bbox{x}]=\omega [z]=\prod _{r}\omega (z_{r})$.  In this case we 
can easily make a semiclassical analysis of the one-point function 
$\langle \varphi (\bbox{x})\rangle =\omega (z)$ by setting
\begin{equation}
e^{-\a(\bbox{x})}e^{i\chi (\bbox{x})} \sqrt {\rho _{0}+\varrho  (\bbox{x})} = \omega (z) ,
\label{ClassGenerA}
\end{equation}
based on the parametrization (\ref{BareDress}).  Here and hereafter, we use the 
same symbols $\a(\bbox{x})$, $\varrho  (\bbox{x})$ and $\chi (\bbox{x})$ also for the classical fields.  When 
an analytic function $\omega (z)$ is given, (\ref{ClassGenerA}) is an integral equation 
determining the density deviation $\varrho  (\bbox{x})$ confined within the \LLL. 

We transform (\ref{ClassGenerA}) into a differential equation.  The 
Cauchy-Riemann equation for the analytic function (\ref{ClassGenerA}) yields,
\begin{equation}
\partial _{j}\bigl(\a(\bbox{x}) - \ln \sqrt {\rho _{0}+\varrho  (\bbox{x})}\bigr) = -\varepsilon _{jk}\partial _{k}\chi (\bbox{x}) .
\label{StepB}
\end{equation}
Using (\ref{StepA}) we obtain that
\begin{equation}
{\nu \over 4\pi }\bbox{\nabla }^{2}\ln\Bigl(1+{\varrho  (\bbox{x})\over \rho _{0}}\Bigr) - \varrho  (\bbox{x}) = \nu Q^{V}_{0}(\bbox{x}) ,
\label{SolitEqVorte}
\end{equation}
which we call the soliton equation, where
\begin{equation}
Q^{V}_{0}(\bbox{x}) = {1\over 2\pi }\varepsilon _{jk}\partial _{j}\partial _{k}\chi (\bbox{x}) = {1\over 2\pi i}\varepsilon _{jk}\partial _{j}\partial _{k}\ln \omega (z) 
\label{VorteCharg}
\end{equation}
is the topological charge density associated with the excitation.  This is 
nonvanishing since $\ln\omega (z)$ is a multivalued function unless 
$\omega (z)=\text{constant}$.  The topological current is
\begin{equation}
Q_{\mu }^{V}(\bbox{x}) = {1\over 2\pi }\varepsilon _{\mu \nu \lambda }\partial _{\nu }\partial _{\lambda }\chi (\bbox{x}) ,
\end{equation}
which is a conserved quantitity, $\partial ^{\mu }Q_{\mu }^{V}(\bbox{x})=0$.  Topological solitons are 
generated around the zeros of $\omega (z)$ according to the soliton equation 
(\ref{SolitEqVorte}).  The density modulation is induced in order to confine the 
excitation within the \LLL.

The topological charge is evaluated as
\begin{equation}
Q^{V} = \int \dx Q^{V}_{0}(\bbox{x}) = {1\over 2\pi i}\oint dx_{k} \partial _{k}\ln \omega (z) ,
\label{VorteChargA}
\end{equation}
where the loop integration $\oint $ is made to encircle the excitation at infinity 
($|\bbox{x}|\rightarrow \infty $) provided $\omega (z)$ is regular everywhere.  The topological charge is 
uniquely determined by the asymptotic behavior of the classical field $\varphi (\bbox{x})$,
\begin{equation}
\varphi (\bbox{x}) \rightarrow  \sqrt {\rho _{0}}z^{q} ,\quad \quad \text{as}\quad  |z| \rightarrow  \infty  ,
\label{VorteAsymp}
\end{equation}
for which the electron number is
\begin{equation}
\Delta N = \int \dx \varrho  (\bbox{x}) = -\nu  Q_{\text{V}} = -\nu q ,
\label{ElectNumbe}
\end{equation}
as follows from the soliton equation (\ref{SolitEqVorte}).  The electric charge 
carried by this soliton is $-e\Delta N=\nu qe$.  It is a hole made in the condensate of 
composite bosons, as is illustrated for the $q=1$ vortex in 
Fig.\ref{FQHE1PS}(c).  We may as well derive the electric number (\ref{ElectNumbe}) 
more directly from the parametrization (\ref{ClassGenerA}).  We take the asymptotic 
behavior $|z|\rightarrow \infty $ in (\ref{ClassGenerA}) and equate it with (\ref{VorteAsymp}),
\begin{equation}
\varrho  (\bbox{x})\rightarrow 0,\quad \quad  \chi (\bbox{x})\rightarrow q\theta , \quad \quad  \a(\bbox{x})\rightarrow -q\ln|z|, \quad \quad \text{as}\quad  |z| \rightarrow  \infty  .
\label{AsympChara}
\end{equation}
Taking the limit $|\bbox{x}|\rightarrow \infty $ in (\ref{SpinB}) we recover the quantization of the 
electron number (\ref{ElectNumbe}).  Actually we have determined the coefficient 
$\sqrt {\rho _{0}}$ in the asymptotic behavior (\ref{VorteAsymp}) in this way.

A topological excitation carries a quantized topological charge, which 
has to be created all at once.  It cannot be created by an accumulation of 
perturbative fluctuations, as agrees with the perturbative result mentioned at 
the begining of this section.  In the absence of the Coulomb term all these 
excitations are degenerate with the ground state, which explains the 
degeneracy in the lowest Landau level at $\nu <1$.  The degeneracy is removed 
since any density moduration acquires a Coulomb energy,
\begin{equation}
\langle H_{C}\rangle  = {e^{2}\over 2\varepsilon }\int \dx\dy {\varrho  (\bbox{x})\varrho  (\bbox{y})\over |\bbox{x}-\bbox{y}|} = \alpha _{C}\nu ^{2}q^{2}{e^{2}\over \varepsilon \ell _{B}} ,
\label{CouloEnerg}
\end{equation}
where $\alpha _{C}$ is a constant of order one.

An explicit example is given by a vortex (quasihole) sitting at $\bbox{x}=0$, 
whose wave function is $\F_{\varphi }[\bbox{x}]=\prod _{r}z_{r}$ up to a normalization factor, or 
$\omega (z)=\sqrt {\rho _{0}}z$.  In this example the topological charge (\ref{VorteCharg}) is 
concentrated at the vortex center, $Q^{V}_{0}(\bbox{x})=\delta (\bbox{x})$.  A crude approximation of 
the soliton equation (\ref{SolitEqVorte}) reads
\begin{equation}
{\ell _{B}^{2}\over 2}\bbox{\nabla }^{2}\varrho  (\bbox{x}) - \varrho  (\bbox{x}) = \nu \delta (\bbox{x}) ,
\end{equation}
since $\nu =2\pi \rho _{0}\ell _{B}^{2}$.  Its exact solution \cite{LaughlinA} is $\varrho  (\bbox{x})=-(\nu /\pi \ell _{B}^{2})K_{0}(s)$ 
with $s=\sqrt {2}r/\ell _{B}$ and $K_{0}(s)$ the modified Bessel function.  This is a rather 
poor approximation because of its singular behavior, $\varrho  (\bbox{x})\rightarrow -\infty $, at the vortex 
center ($\bbox{x}=0$).  A better approximation is given by
\begin{equation}
\varrho  (\bbox{x}) = -\rho _{0}\Bigl(1+s-{s^{2}\over 6}\Bigr)e^{-s} ,
\label{BetteAppro}
\end{equation}
which has the correct behavior both at $\bbox{x}=0$ and $|\bbox{x}|\gg \ell _{B}$.  Furthermore, it 
has the correct topological charge.  For a numerical analysis, it is 
convenient to set $\rho (\bbox{x})=\rho _{0}e^{u(s)}$ in the soliton equation (\ref{SolitEqVorte}), as 
yields
\begin{equation}
{d^{2}u\over ds^{2}} + {1\over s}{du\over ds} + 1 = e^{u(s)} ,
\label{ModifLiuvi}
\end{equation}
for $s>0$.  The result of a numerical analysis shows that the density 
modulation is well approximated by (\ref{BetteAppro}), as in Fig.\ref{VortNumePS}.  
The Coulomb energy is given by (\ref{CouloEnerg}) with $\alpha _{C}\simeq 0.39$.
\FigOut{\FigVortNume}
\subsection{Antivortices}

We have so far considered the case where the $N$-point function is 
factorizable, $\omega [z]=\prod _{r}\omega (z_{r})$.  A vortex is described by the choice of 
$\omega [z]=\prod _{r}z_{r}$ up to a normalization, which gives an angular momentum to each 
electron.  Similarly, an antivortex is generated by decreasing an angular 
momentum from each electron.  Naively, this is to multiply $z_{r}^{*}$ to the $r$th 
electron, but we cannot accept this option because the resulting configuration 
does not stay within the lowest Landau level.  The only possible way seems to 
use the derivative operation $\prod \partial /\partial z_{r}$ to decrease the angular 
momentum \cite{LaughlinA}.  It is convenient to write symbolically as
\begin{equation}
\omega [z] = \prod _{r}{\partial \over \partial z_{r}} ,
\label{AntiVOmega}
\end{equation}
up to a normalization, with the understanding that it acts only on the 
polynomial part of the wave function (\ref{WaveElect}).  Then, the wave function 
$\F_{\varphi }[\bbox{x}]$ remains to be analytic but becomes very complicated.  It is no longer 
factorizable, as makes the analysis of antivortices considerably complicated.

We analyze a generic excitation on the Laughlin state.  We propose to 
define one-point function by the formula,
\begin{equation}
\langle \varphi (\bbox{x})\rangle  = \lim_{N\rightarrow \infty }\langle \F^{N}|\varphi (\bbox{x})|\F^{N+1}\rangle  , \label{OneNwave}
\end{equation}
by taking two states $|\F^{N}\rangle $ and $|\F^{N+1}\rangle $ which are identical except the 
number of electrons in the ground state.  Let us give an example of a single 
vortex sitting at the center of a large system.  Then, these two states are 
defined by the wave functions $\F^{N}[\bbox{x}]=\prod _{r=1}^{N}z_{r}\F^{N}_{\text{LN}}[\bbox{x}]$ and 
$\F^{N+1}[\bbox{x}]=\prod _{r=1}^{N+1}z_{r}\F^{N+1}_{\text{LN}}[\bbox{x}]$, where $\F^{N}_{\text{LN}}[\bbox{x}]$ is the Laughlin 
wave function containing $N$ electrons.  The semiclassical property of the 
vortex is independent of the number $N$ of electrons in the system provided 
$N$ is large.  

Using (\ref{NbodyRb}) and (\ref{NbodyRc}) in (\ref{OneNwave}), we may express $\langle \varphi (\bbox{x})\rangle $ 
as
\begin{equation}
\langle \varphi (\bbox{x})\rangle  = \int [d\bbox{x}] \omega (z_{1},z_{2},\cdots ,z_{N})^{*}\omega (z_{1},z_{2},\cdots ,z_{N},z)
|\F^{N}_{\text{LN}}[\bbox{x}]|^{2} .
\label{GenerOneBody}
\end{equation}
This is not an analytic function in general, although we recover $\langle \varphi (\bbox{x})\rangle =\omega (z)$ 
when factorizable, $\omega [z]=\prod _{r}\omega (z_{r})$.  Nevertheless, by inspecting the 
integration (\ref{GenerOneBody}), we may conclude that it becomes an analytic 
function asymptotically, $\langle \varphi (\bbox{x})\rangle \rightarrow \omega (z)$ as $|z|\rightarrow \infty $, since the integration over 
$\bbox{x}_{r}$ is convergent due to the gaussian factor in the Laughlin function.  See 
(\ref{ExampA}) and (\ref{ExampB}) explicitly.

Although the Cauchy-Rieman equation (\ref{StepB}) does not follow, we obtain 
the soliton equation (\ref{SolitEqVorte}) with
\begin{equation}
Q_{0}^{V}(\bbox{x}) = {1\over 2\pi }\bbox{\nabla }^{2}(\ln \langle \varphi \rangle  - i\chi ) .
\end{equation}
In evaluating the electron number $\int d^{2}x\varrho  (\bbox{x})$, therefore, only the bounday 
value of $\langle \varphi (\bbox{x})\rangle $ is relevant, where it approaches an analytic function 
$\omega (z)$.  The electron number is given by the same formula as (\ref{ElectNumbe}) in 
terms of the asymptotic behavior of $\langle \varphi (\bbox{x})\rangle $.

We give an example for the antivortex (quasielectron).  Evaluating the 
polynomial part in (\ref{WaveElect}) with (\ref{AntiVOmega}) explicitly, the $N$-body 
wave function $\omega _{qe}[z]$ for an antivortex sitting at the origin is given 
by way of the formula \cite{LaughlinA}
\begin{equation}
\prod _{t}{\partial \over \partial z_{t}} \prod _{r<s} (z_{r}-z_{s})^{m} \equiv  \omega _{qe}[z] \prod _{r<s} (z_{r}-z_{s})^{m} ,
\end{equation}
or 
\begin{equation}
\F^{qe}_{\varphi }[z] = \omega _{qe}[z] = {\prod _{t}{\partial \over \partial z_{t}} \prod _{r<s} (z_{r}-z_{s})^{m}\over \prod _{r<s} (z_{r}-z_{s})^{m}} .
\label{AntiVorteX}
\end{equation}
This is the wave function of a quasielectron in terms of composite bosons.  
The one-point function (\ref{GenerOneBody}) is
\begin{equation}
\langle \varphi _{qe}(\bbox{x})\rangle  = \int [d\bbox{x}] \omega _{qe}(z_{1},z_{2},\cdots ,z_{N})^{*}\omega _{qe}(z_{1},z_{2},\cdots ,z_{N},z)|\F^{N}_{\text{LN}}[\bbox{x}]|^{2} .
\label{OneBodyAntiv}
\end{equation}
The dominant term for $|z|\gg 1$ is given by
\begin{equation}
\langle \varphi _{qe}(\bbox{x})\rangle  \simeq  G(\bbox{x}) = \sum _{s=1}^{N}\int [d\bbox{x}]{1\over z-z_{s}}|\F_{qe}^{N}(\bbox{x}_{1},\cdots ,\bbox{x}_{N})|^{2} ,
\label{ExampA}
\end{equation}
up to a normalization factor, where $\F_{qe}^{N}=\omega _{qe}\F_{\text{LN}}^{N}$ is the 
$N$-body quasielectron wave function.  We calculate $G(\bbox{x})$ in the domain 
$|\bbox{x}|\gg 1$, where we may expand $(z-z_{s})^{-1}$ in a power series of $z_{s}/z$, and find 
$G(\bbox{x})=z^{-1}$.  If it were an analytic function, we would have $G(\bbox{x})=z^{-1}$ by 
analytic continuation.  However, since $G(0)=0$ at $z=0$ by integrating over 
the angle variable, it cannot be so, though it approaches an analytic function 
asymptotically.

We give an explicit example of the function of type (\ref{ExampA}),
\begin{equation}
G_{1}(\bbox{x})={1\over \pi }\int {d^{2}\omega \over z-\omega } e^{-|\omega |^2} = {1\over z}\Bigl(1-e^{-|z|^2}\Bigr) .
\label{ExampB}
\end{equation}
We explicitly see that $G_{1}(\bbox{x}) \rightarrow  z^{-1}$ exponentially fast for $|z|\gg 1$ and yet 
$G_{1}(\bbox{x})=0$ at $\bbox{x}=0$.

We may in principle determine the classical configuration of the 
antivortex from (\ref{BareDress}) once we obtain a closed formula for 
(\ref{OneBodyAntiv}) in the limit $N\rightarrow \infty $, which is yet to be done.  However, it is 
clear that $\langle \varphi (\bbox{x})\rangle $ approaches its asymptotic value $z^{-1}$ exponentially fast 
outside the core of size $\sim \ell _{B}$.  The antivortex is a lump of a quantized 
electric charge $-e/m$ in the homogeneous ground state as illustrated in 
Fig.\ref{FQHE1PS}(d).  We expect that the activation energy of one antivortex 
is nearly the same as that of one vortex.  In general we have the asymptotic 
behavior (\ref{VorteAsymp}) for vortex ($q>0$) and antivortex ($q<0$) excitations.  
It characterizes topological excitations on the Laughlin state, as we have 
noticed in (\ref{AsympChara}).  Thirmal fluctuations generate vortex-antivortex 
pairs as in Fig.\ref{FQHE1PS}(e), which is ditected by measuring the 
longitudinal resistivity (\ref{LongiResis}).
\section{Quantum Hall Ferromagnet}\label{SecQHFerro}

We proceed to analyze the QH system with the SU(2) symmetry.  The 
electron field $\psi ^{\alpha }(\bbox{x})$ has the index $\alpha =\upA ,\dnA $.  It denotes the electron spin in 
the monolayer QH system with the spin SU(2) symmetry, or the layer index in a 
certain bilayer QH system with the pseudospin SU(2) symmetry.  For 
definiteness we analyze the monolayer spin system in this paper.

The Hamiltonian depends on the electron spin through the Zeeman energy 
term,
\begin{equation}
H_{Z} = -g^{*}\mu _{B}B \int d^{2}x S^{z}(\bbox{x}) ,
\label{ZeemaTerm}
\end{equation}
with $S^{z}={1\over 2}(\psi ^{\upA \dagger }\psi ^{\upA }-\psi ^{\dnA \dagger }\psi ^{\dnA })$, where $g^{*}$ is the gyromagnetic factor and $\mu _{B}$ 
the Bohr magneton.  Each Landau level contains two energy levels with the 
one-particle gap energy $g^{*}\mu _{B}B$.  The lowest Landau level is filled at $\nu =2$.  
We consider the case where the Zeeman energy is much smaller than the Coulomb 
energy.  Though one Landau level contains two degenerate energy levels in the 
vanishing limit of the Zeeman coupling ($g^{*}=0$), the system becomes 
incompressible at $\nu =1/m$.  The physical reason is the Coulomb exchange 
energy, as we now discuss.

We define the \textit{bare} composite-boson field $\phi ^{\alpha }(\bbox{x})$ and the 
\textit{dressed} composite-boson field $\varphi ^{\alpha }(\bbox{x})$ by 
\begin{equation}
\phi ^{\alpha }(\bbox{x}) = e^{-i\Theta (\bbox{x})}\psi ^{\alpha }(\bbox{x}),\quad  \varphi ^{\alpha }(\bbox{x}) = e^{-\a(\bbox{x})}\phi ^{\alpha }(\bbox{x}),
\end{equation}
where the phase field $\Theta (\bbox{x})$ and the auxiliary field $\a(\bbox{x})$ are given by 
(\ref{PhaseTheta}) and (\ref{SpinB}), respectively, with the total electron density 
$\rho (\bbox{x})=\sum _{\alpha }\psi ^{\alpha \dagger }(\bbox{x})\psi ^{\alpha }(\bbox{x})=\sum _{\alpha }\varphi ^{\alpha \dagger }(\bbox{x})e^{2\a(\bbox{x})}\varphi ^{\alpha }(\bbox{x})$.  The Hamiltonian is
\begin{align}
H &={1\over 2M}\sum _{\alpha }\int \dx \psi ^{\alpha \dagger }(\bbox{x})(P_{x}^{2}+P_{y}^{2})\psi ^{\alpha }(\bbox{x}) + H_{C}+H_{Z}  \label{HamilSpinA}\\
&= \omega _{c }\sum _{\alpha }\int \dx \Bigl({\partial \over \partial z^{*}}\varphi ^{\alpha }(\bbox{x})\Bigr)^{\dagger }e^{2\a(\bbox{x})}{\partial \over \partial z^{*}}\varphi ^{\alpha }(\bbox{x}) +{N\over 2}\omega _{c}+H_{C}+H_{Z} ,
\label{HamilSpinB}
\end{align}
with the Coulomb term $H_{C}$ and the Zeeman term $H_{Z}$.  The Coulomb term depends 
on the deviation $\varrho  (\bbox{x})$ of the total electron density from the average 
density, $\varrho  (\bbox{x})=\rho (\bbox{x})-\rho _{0}$, as in (\ref{CouloEnergPre}).
\subsection{Spin Texture}

We may decompose the bare \CB field into the U(1) field $\phi (\bbox{x})$ and the 
SU(2) field $n^{\alpha }(\bbox{x})$, 
\begin{equation}
\phi ^{\alpha }(\bbox{x}) = \phi (\bbox{x})n^{\alpha }(\bbox{x}), \quad \sum _{\alpha }n^{\alpha \dagger }n^{\alpha }= 1 .
\label{BareSpin}
\end{equation}
The field $n^{\alpha }(\bbox{x})$ is the complex-projective (\CP) field\cite{Skyrmion}, whose 
overall phase has been removed and given to the U(1) field $\phi (\bbox{x})$.  The spin 
operator is expressed as 
\begin{equation}
S^{a}(\bbox{x})={1\over 2}\rho (\bbox{x})\Sigma ^{a}(\bbox{x}) ,
\end{equation}
where
\begin{equation}
\Sigma ^{a}(\bbox{x})=\bbox{n}^{\dagger }(\bbox{x}){\tau ^{a}}\bbox{n}(\bbox{x}), \quad  \bbox{n}(\bbox{x}) = \begin{pmatrix}n^{\upA }(\bbox{x})\\n^{\dnA }(\bbox{x})\end{pmatrix}.
\end{equation}
In terms of the dressed composite-boson field the decomposition reads
\begin{equation}
\varphi ^{\alpha }(\bbox{x}) = \varphi (\bbox{x})n^{\alpha }(\bbox{x}), \quad  \varphi (\bbox{x}) = e^{-\a(\bbox{x})}\phi (\bbox{x}) .
\label{SpinAD}
\end{equation}
The SU(2) component $n^{\alpha }(\bbox{x})$ is common between the bare and dressed fields 
(\ref{BareSpin}) and (\ref{SpinAD}):  It is a local field.  On the other hand, $\varphi (\bbox{x})$ is 
a nonlocal field due to the factor $e^{-\a(\bbox{x})}$ as in the spinless theory.

The ground state minimizes both the Coulomb and Zeeman energies.  The 
Coulomb energy is minimized by the homogeneous electron density, $\langle \rho (\bbox{x})\rangle =\rho _{0}$.  
The Zeeman energy is minimized when all electrons are polarized into the 
positive $z$ axis, $\langle n^{\upA }(\bbox{x})\rangle =1$ and $\langle n^{\dnA }(\bbox{x})\rangle =0$.  The ground state is unique, 
which we denote by $|g_{0}\rangle $.

We consider a spin texture given by performing an SU(2) transformation 
on the ground state $|g_{0}\rangle $,  $|\FF\rangle =e^{i\O}|g_{0}\rangle $, where $\O$ is its generator,
\begin{equation}
\O = \sum _{a}\int \dx f^{a}(\bbox{x}) S^{a}(\bbox{x}) = \sum _{a}\int {\dq}f^{a}_{-\bbox{q}}S^{a}_{\bbox{q}}  .
\label{GenerSU}
\end{equation}
It is described by the classical sigma field $s^{a}(\bbox{x})=\langle \FF|\Sigma ^{a}(\bbox{x})|\FF\rangle $, which we 
parametrize as
\begin{align}
s^{x}(\bbox{x})=\sigma (\bbox{x}),  \quad  s^{y}(\bbox{x})= \sqrt {1-\sigma ^{2}(\bbox{x})}\sin\vartheta (\bbox{x}), \quad  s^{z}(\bbox{x})=\sqrt {1-\sigma ^{2}(\bbox{x})}\cos\vartheta (\bbox{x}). 
\label{ClassPS}
\end{align}
The spin texture is classified by the Pontryagin number \cite{Skyrmion}, 
$Q^{P}=\int \dx Q^{P}_{0}(\bbox{x})$, with the topological current
\begin{equation}
Q_{\mu }^{P}(\bbox{x}) = {1\over 8\pi }\varepsilon _{abc}\varepsilon _{\mu \nu \lambda }s_{a}\partial ^{\nu }s_{b}\partial ^{\lambda }s_{c}  .
\label{PontrNumbe}
\end{equation}
It is absolutely conserved, $\partial ^{\mu }Q_{\mu }^{P}=0$.  We note that the spin texture $|\FF\rangle $ 
does not belong to the \LLL.
\subsection{Wave Functions}

The two-component composite-boson field is $\Phi (\bbox{x})=\varphi (\bbox{x})\bbox{n}(\bbox{x})$.  With the 
Hamiltonian (\ref{HamilSpinB}), the LLL condition for the state $|\F\rangle $ is
\begin{equation}
{\partial \over \partial z^{*}}\Phi (\bbox{x})|\F\rangle  = 0 .
\label{LLLcondiBL}
\end{equation}
Because of this condition the $N$-body wave function is analytic,
\begin{equation}
\F_{\varphi }[\bbox{x}] = \langle 0|\Phi (\bbox{x}_{1})\Phi (\bbox{x}_{2}) \cdots  \Phi (\bbox{x}_{N})|\F\rangle  = \Omega [z] ,
\label{DressWaveFunct}
\end{equation}
where $\Omega [z]$ is totally symmetric in $N$ variables.  When it is factorizable, 
$\Omega [z]=\prod _{r}\Omega (z_{r})$, it has a simple expression,
\begin{equation}
\Omega (z) = \begin{pmatrix}\omega ^{\upA }(z)\\ \omega ^{\dnA }(z)\end{pmatrix} .
\end{equation}
For the ground state $|g_{0}\rangle $ we have
\begin{equation}
\Omega (z) = \sqrt {\rho _{0}}\begin{pmatrix}1\\ 0\end{pmatrix}. 
\label{VacuuMono}
\end{equation}
In terms of the original electrons the wave function is given by 
\begin{equation}
\Psi _{\varphi }[z,z^{*}]=\Omega [z]\Laugh ,
\end{equation}
with the Laughlin wave function $\Laugh$.  
\subsection{LLL Projection}

The energy of the spin texture $\langle \FF|H|\FF\rangle $ acquires a contribution 
from the kinetic Hamiltonian, which is of the order of the Landau-level gap 
energy $\hbar \omega _{c}$.  This is because the spin texture $|\FF\rangle $ does not belong to the 
\LLL.  It is necessary to excite only the component $|\F\rangle $ belonging to the 
\LLL by requiring (\ref{LLLcondiBL}).  We have two complementary methods.  One 
method is to solve the LLL condition (\ref{LLLcondiBL}) semiclassically and 
determine the density modulation as well as the spin modulation, as is the 
method we have used to analyze the vortex excitation in the previous section.  
We use the same method to analyze the Skyrmion excitation in the following 
section.  The other is an algebraic method based on the LLL projection 
\cite{refLLL}.  Here, we use it to derive the effective Hamiltonian govering a 
small spin fluctuation around the ground state.

The LLL component $|\F\rangle $ is extracted from the spin texture $|\FF\rangle $ by 
extracting the LLL component $\llangle f^{a}(\bbox{x})\rrangle $ from the "wave packet" $f^{a}(\bbox{x})$ in the 
generator $\O$ of the SU(2) transformation (\ref{GenerSU}).  This turns out to 
replace the plane weve $e^{i\bbox{x}\bbox{q}}$ with \cite{EzaIQC}
\begin{equation}
\llangle e^{i\bbox{x}\bbox{q}}\rrangle \equiv e^{-{1\over 4}\bbox{q}^{2}\ell _{B}^{2}}e^{i\bbox{X}\bbox{q}} 
\end{equation}
in the Fourier representation of $f^{a}(\bbox{x})$, where $\bbox{X}=(X,Y)$ is the guiding 
center.  We call $\llangle e^{i\bbox{x}\bbox{q}}\rrangle $ the LLL projection of $e^{i\bbox{x}\bbox{q}}$.  The generator 
(\ref{GenerSU}) is projected as
\begin{equation}
\OO = \sum _{a}\int \dx \llangle f^{a}(\bbox{x})\rrangle  S^{a}(\bbox{x}) = \sum _{a}\int {\dq}f^{a}_{-\bbox{q}}\widehat{S}^{a}_{\bbox{q}}  ,
\end{equation}
where
\begin{equation}
\widehat{S}^{a}_{\bbox{q}}=(2\pi )^{-1}\int \dx \llangle e^{-i\bbox{q}\bbox{x}}\rrangle S^{a}(\bbox{x}).
\end{equation}
Similarly we define
\begin{equation}
\widehat{\rho }_{\bbox{q}}=(2\pi )^{-1}\int \dx \llangle e^{-i\bbox{q}\bbox{x}}\rrangle \rho (\bbox{x}) 
\end{equation}
for the electron density operator.  

From the commutation relation $[X,Y]=-i\ell _{B}^{2}$ between the $X$ and $Y$ 
components of the guiding center, we obtain
\begin{equation}
[e^{i\bbox{q}\bbox{X}}, e^{i\bbox{p}\bbox{X}}] = 2ie^{i(\bbox{q}+\bbox{p})\bbox{X}} \sin\bigl[\ell _{B}^{2}{\bbox{q}\!\wedge\!\bbox{p}\over 2}\bigr], \label{Walgebra}
\end{equation}
with $\bbox{q}\!\wedge\!\bbox{p}\equiv q_{x}p_{y}-q_{y}p_{x}$.  The translation $e^{i\bbox{q}\bbox{x}}$ is Abelian, but the magnetic 
translation $e^{i\bbox{q}\bbox{X}}$ is non-Abelian.  It governs the symmetric structure of the 
two-dimensional space after the LLL projection.  It is straightforward to 
derive the following W$_{\infty }\times $SU(2) algebra \cite{EzaIQC},
\begin{subequations}\label{SUCommu}
\begin{align}
&[\widehat{\rho }_{\bbox{p}},\widehat{\rho }_{\bbox{q}}]={i\over \pi }\widehat{\rho }_{\bbox{p}+\bbox{q}}\sin\bigl[\Wed{\bbox{p}}{\bbox{q}}\bigr]\exp\bigl[{\ell _{B}^{2}\over 2}\bbox{p}\bbox{q}\bigr], \label{SUCommuA}\\
&[\widehat{S}^{a}_{\bbox{p}},\widehat{\rho }_{\bbox{q}}]={i\over \pi }\widehat{S}^{a}_{\bbox{p}+\bbox{q}}\sin\bigl[\Wed{\bbox{p}}{\bbox{q}}\bigr]\exp\bigl[{\ell _{B}^{2}\over 2}\bbox{p}\bbox{q}\bigr],\label{SUCommuB}\\
&[\widehat{S}^{a}_{\bbox{p}}, \widehat{S}^{b}_{\bbox{q}}]={i\over 2\pi }\varepsilon ^{abc}\widehat{S}^{c}_{\bbox{p}+\bbox{q}}\cos\bigl[\Wed{\bbox{p}}{\bbox{q}}\bigr]\exp\bigl[{\ell _{B}^{2}\over 2}\bbox{p}\bbox{q}\bigr] \notag\\
&\hspace*{20mm}+{i\over 4\pi }\delta ^{ab}\widehat{\rho }_{\bbox{p}+\bbox{q}}\sin\bigl[\Wed{\bbox{p}}{\bbox{q}}\bigr]\exp\bigl[{\ell _{B}^{2}\over 2}\bbox{p}\bbox{q}\bigr],
\label{SUCommuC}
\end{align}
\end{subequations}
based on the algebra (\ref{Walgebra}) of the magnetic translation.

We evaluate the energy $\langle \F|H|\F\rangle $ by making a perturbative expansion 
of the spin texture around the ground state,
\begin{equation}
H_{\text{eff}}\equiv \langle \F|H|\F\rangle  = \langle g_{0}|H|g_{0}\rangle  - \langle g_{0}|[\OO, H]|g_{0}\rangle  + \cdots  .
\end{equation}
The Hamiltonian $H$ contains the physical degree of freedom associated with 
both the relative coordinate $\bbox{R}$ and the guiding center $\bbox{X}$.  Since they 
commute each other, the relative coordinate $\bbox{R}$ is easily eliminated by 
operating it to the ground state $|g_{0}\rangle $ or $\langle g_{0}|$.  The result is
\begin{equation}
H_{\text{eff}}= \langle g_{0}|\widehat{H}|g_{0}\rangle  - \langle g_{0}|[\OO, \widehat{H}]|g_{0}\rangle  + \cdots  ,
\label{AnalsStepA}
\end{equation}
where $\widehat{H}$ is the LLL-projected Hamiltonian.  For instance, the Coulomb energy 
is projected as
\begin{equation}
\widehat{H}_{C} = {e^{2}\over 2\varepsilon }\int \dx\dy \varrho  (\bbox{x})\llangle V(\bbox{x}-\bbox{y})\rrangle \varrho  (\bbox{y}) = \pi  \int {\dq}V(\bbox{q}) \widehat{\rho }_{-\bbox{q}}\widehat{\rho }_{\bbox{q}} ,
\end{equation}
where $V(\bbox{q})$ is the Fourier transformation of the potential $V(\bbox{x})=1/|\bbox{x}|$.  
Making a straightforward algebraic calcutaion in (\ref{AnalsStepA}), making a 
gradient expansion and taking the lowest order term in $f^{a}_{-\bbox{q}}$, we obtain 
\cite{EzaIQC,MG}
\begin{equation}
H_{\text{eff}}= {1\over 2}\rho _{s}\sum _{a}\int \dx [\partial _{k}s^{a}(\bbox{x})]^{2} - {\rho _{0}\over 2}g^{*}\mu _{B}B \int \dx s^{z}(\bbox{x}) ,
\label{EnergChangSPN}
\end{equation}
in terms of the sigma field.  It is consistent with another perturbative 
result found in an earlier reference \cite{KallinHalperin}.  The first term 
represents the spin stiffness due to the Coulomb exchange energy,  
$\rho _{s}=\nu e^{2}/(16\sqrt {2\pi }\varepsilon \ell _{B})$.  It arises because of the following reason:  The local 
spin rotation has components in higher Landau level since it is not a symmetry 
of the Hamiltonian (\ref{HamilSpinA}).  Only its LLL component is excited at 
sufficiently low temperature.  Since the LLL components of the spin operators 
and the density operator do not commute as in (\ref{SUCommuB}), the local spin 
rotation induces a local density modulation and affects the Coulomb energy.  
\subsection{Goldstone Mode}

The Zeeman effect is quite small in actual samples.  We consider the 
vanishing limit of the Zeeman term ($g^{*}=0$).  According to the effective 
Hamiltonian (\ref{EnergChangSPN}), the energy is minimized for any constant value 
of the sigma field, $\bbox{s}(\bbox{x})=\bbox{s}_{0}=$constant.  Hence, there exists a degeneracy in 
the ground states as indexed by $\bbox{s}_{0}$.  The choice of a ground state implies a 
spontaneous magnetization, or a \textit{quantum Hall ferromagnetism}.  When a 
continuous symmetry is spontaneously broken, there should arise a gapless mode 
known as the Goldstone mode.  Quantum coherence develops spontaneously.  

Actually, there is a small Zeeman effect in actual samples, and the 
ground state $|g_{0}\rangle $ is chosen where $\bbox{s}_{0}=(0,0,1)$.  However, we can treat the 
Zeeman interaction as a perturbation because it is less important than the 
Coulomb interaction.  The effective Hamiltonian (\ref{EnergChangSPN}) is reliable 
also in the presence of the Zeeman interaction ($g^{*}\not=0$), though the 
Goldstone mode is no longer gapless.  The key property of the QH ferromagnet 
is that it is a coherent state, where all spin components $S^{x}(\bbox{x})$, $S^{y}(\bbox{x})$ and 
$S^{z}(\bbox{x})$ are simultaneously observable, $s^{a}(\bbox{x})=2\rho _{0}^{-1}\langle S^{a}(\bbox{x})\rangle $.  An evidence is 
the existence of coherent excitations such as Skyrmions.

The Goldstone mode describes small fluctuations of the \CP field around 
the ground state (\ref{VacuuMono}).  Up to the lowest order of the perturbation in 
the \CP field $\bbox{n}(\bbox{x})$, it is parametrized as \cite{EzaIQC}
\begin{equation}
n^{\upA }(\bbox{x})=1, \quad \quad  n^{\dnA }(\bbox{x})={\zeta (\bbox{x})\over \sqrt {\rho _{0}}},
\label{CPGolds}
\end{equation}
with $[\zeta ^{\dagger }(\bbox{x}),\zeta (\bbox{y})]=i\delta (\bbox{x}-\bbox{y})$.  The LLL condition (\ref{LLLcondiBL}) yields two 
conditions,
\begin{equation}
{\partial \over \partial z^{*}} \varphi (\bbox{x})|\F\rangle  = 0 ,\quad  {\partial \over \partial z^{*}} \zeta (\bbox{x})|\F\rangle  = 0 ,
\end{equation}
up to this order.  Although they look similar, they describe very different 
excitation modes.  As in the spinless QH system, $\varphi (\bbox{x})$ is a nonlocal field 
and generates extended objects.  On the other hand, $\zeta (\bbox{x})$ is a local field, 
and it describes the Goldstone mode.

We may relate $\zeta (\bbox{x})$ to the classical sigma field (\ref{ClassPS}),
\begin{equation}
\langle \zeta (\bbox{x})\rangle  = {\sqrt {\rho _{0}}\over 2}\bigl\{\sigma (\bbox{x})+i\vartheta (\bbox{x})\bigr\} .
\label{GoldsPertu}
\end{equation}
The effective Hamiltonian (\ref{EnergChangSPN}) is recognized as a classical 
counterpart of the quantum version,
\begin{equation}
H_{\text{eff}}=  {2\rho _{s}\over \rho _{0}}\int d^{2}x \bigl[\partial _{k}\zeta ^{\dagger }(\bbox{x})\partial _{k}\zeta (\bbox{x}) + \xi _{L}^{-2}\zeta ^{\dagger }(\bbox{x})\zeta (\bbox{x})\bigr],
\end{equation}
on the coherent state.  Here, $\xi _{L}$ is the coherent length,
\begin{equation}
\xi _{L}=\sqrt {2\rho _{s}\over g^{*}\mu _{B}B\rho _{0}} = {(2\pi )^{1/4}\ell _{B}\over 2\sqrt {2\widetilde{g}^{ }} },
\label{CoherLengt}
\end{equation}
where $\widetilde{g}=g^{*}\mu _{B}B/(e^{2}/\varepsilon \ell _{B})$ is the ratio of the Zeeman energy to the Coulomb 
energy.  We have $\xi _{L}\sim 4\ell _{B}$ in typical samples at $B\simeq 10$ Tesla, where $\widetilde{g}\simeq 0.02$.  
In the momentum space the effective Hamiltonian reads
\begin{equation}
\H_{\rm{eff}}(\bbox{k}) = E_{\bbox{k}}\zeta ^{\dagger }_{\bbox{k}}\zeta _{\bbox{k}} ,
\end{equation}
with $[\zeta _{\bbox{k}}, \zeta ^{\dagger }_{\bbox{l}}]=\delta (\bbox{k}-\bbox{l})$, and the dispersion relation is
\begin{equation}
E_{\bbox{k}} = {2\rho _{s}\over \rho _{0}}\bbox{k}^{2} + g^{*}\mu _{B}B.
\label{SuperModeZeema}
\end{equation}
The Goldstone mode has acquired a gap $E_{0}=g^{*}\mu _{B}B$.
\section{Topological Excitations}\label{SecSpinTE}

We analyze topological (nonperturbative) excitations on the QH 
ferromagnet.  We use the semiclassical approximation because the algebraic 
analysis is so difficult without making a perturbative expansion.  When the 
$N$-body wave function is factorizable, the one-point function is analytic, 
$\langle \varphi ^{\alpha }(\bbox{x})\rangle =\omega ^{\alpha }(z)$.  From (\ref{SpinAD}), the one-point function is parametrized as
\begin{equation}
e^{-\a(\bbox{x})}e^{i\chi (\bbox{x})} \sqrt {\rho _{0}+\varrho  (\bbox{x})}n^{\alpha }(\bbox{x}) = \omega ^{\alpha }(z) ,
\label{ClassGenerB}
\end{equation}
since $|\phi (\bbox{x})|^{2}=\rho _{0}+\varrho  (\bbox{x})$.  Here and hereafter, all fields are classical fields.  
When the wave function $\omega ^{\alpha }(z)$ is given, the electron density $\varrho  (\bbox{x})$ and the 
spin field $S^{a}(\bbox{x})$ are determined by this equation.  There are two types of 
excitations associated with the U(1) part and the SU(2) part of the \CB field.  
The U(1) excitation has a characteric length $\ell _{B}$, while the SU(2) excitation 
has no scale provided the Zeeman term is neglected.
\subsection{Vortex Excitations}

The U(1) excitation is generated on the spin-polarized ground state 
(\ref{VacuuMono}) when $\partial _{k}\chi (\bbox{x})\not=0$ and $\partial _{k}n^{\alpha }(\bbox{x})=0$ in (\ref{ClassGenerB}).  We may 
set $\langle \varphi ^{\dnA }\rangle =0$.  The one-point function $\langle \varphi ^{\upA }(\bbox{x})\rangle $ is essentially Abelian, and 
the Cauchy-Riemann equation for (\ref{ClassGenerB}) yields precisely the same 
soliton equation (\ref{SolitEqVorte}).  The topological charge density is given by 
(\ref{VorteCharg}) with an analytic function $\omega (z)=\omega ^{\upA }(z)$.  The U(1) excitation is 
the vortex.  The Coulomb energy of the vortex excitation is given by 
(\ref{CouloEnerg}) with $\alpha _{C}\simeq 0.39$.  We should remark that the vortex excitation 
does not induce any spin flip.  There is no antivortex excitation.  Instead of 
it an electron is placed into the spin-down state, as would increase the 
Coulomb energy of the same order as the vortex excitation and the Zeeman 
energy, as illustrated in Fig.\ref{SkyrExciPS}(a).
\subsection{Skyrmion Excitations}

The SU(2) excitation is generated on the spin-polarized ground state 
(\ref{VacuuMono}) when $\partial _{k}\chi (\bbox{x})=0$ and $\partial _{k}n^{\alpha }(\bbox{x})\not=0$ in (\ref{ClassGenerB}).  The \CP 
field is
\begin{equation}
n^{\alpha }(\bbox{x}) = {\omega ^{\alpha }(z)\over \sqrt {|\omega ^{\upA }(z)|^{2}+|\omega ^{\dnA }(z)|^{2}}}, 
\label{GenerSkyrm}
\end{equation}
yielding the wave function
\begin{equation}
\Psi _{\text{Skyrm}}[z,z^{*}] = 
\prod _{r}\begin{pmatrix} \omega ^{\upA }(z_{r})\\ \omega ^{\dnA }(z_{r})\end{pmatrix}_{\kern-1pt r}\Laugh .
\label{SkyrmWave}
\end{equation}
A simplest choice is given by
\begin{equation}
\begin{pmatrix} \omega ^{\upA }(z)\\ \omega ^{\dnA }(z)\end{pmatrix} = \sqrt {\rho _{0}}
\begin{pmatrix} z^{q}\\ (\kappa /2)^{q}\end{pmatrix} ,
\label{SkyrmOmega}
\end{equation}
with a positive integer $q$, which describes a classical Skyrmion with scale 
$\kappa $ sitting at the origin of the system \cite{MG}.  It is clear in (\ref{SkyrmOmega}) 
that the Skyrmion is reduced to the vortex in the limit $\kappa \rightarrow 0$.  We use this 
property to check the consistency of the Skyrmion theory.

The classical sigma field (\ref{ClassPS}) is calculated as
\begin{equation}
s^{x} = \sqrt {1-(s^{z})^{2}}\cos(q\theta ),\quad
s^{y} =-\sqrt {1-(s^{z})^{2}}\sin(q\theta ),\quad
s^{z} = {r^{2q}-(\ell _{B}\kappa )^{2q}\over r^{2q}+(\ell _{B}\kappa )^{2q}},
\label{SkyrmSpin}
\end{equation}
and the Pontryagin number density (\ref{PontrNumbe}) as
\begin{equation}
Q_{0}^{P}(\bbox{x}) = {q^{2}\over \pi } {r^{2q-2}(\ell _{B}\kappa )^{2q}\over [r^{2q}+(\ell _{B}\kappa )^{2q}]^{2}} .
\label{SkyrmPontr}
\end{equation}
The spin flips at the Skyrmion center, $\bbox{s}=(0,0,-1)$ at $r=0$, while the 
spin-polarized ground state is approached away from it, $\bbox{s}=(0,0,1)$ for 
$r\gg \kappa \ell _{B}$.  

The Skyrmion excitation modulates not only the SU(2) part but also the 
U(1) part via the relation (\ref{ClassGenerB}).  The Cauchy-Rieman equation reads
\begin{equation}
\partial _{j}\bigl(\a(\bbox{x}) - \ln \sqrt {\rho _{0}+\varrho  (\bbox{x})}\bigr) = -\varepsilon _{jk}K_{k} ,
\label{SkyrmSPNf}
\end{equation}
where
\begin{equation}
K_{k}= -i \sum _{\alpha  }n^{\alpha *}\partial _{k}n^{\alpha } .
\end{equation}
From (\ref{SkyrmSPNf}) the same soliton equation as (\ref{SolitEqVorte}) is derived,
\begin{equation}
{\nu \over 4\pi }\bbox{\nabla }^{2}\ln\Bigl(1+{\varrho  (\bbox{x})\over \rho _{0}}\Bigr) - \varrho  (\bbox{x}) = \nu Q^{P}_{0}(\bbox{x}) ,
\label{SolitEqSkyrm}
\end{equation}
but the topological charge density now reads
\begin{equation}
Q^{P}_{0}(\bbox{x}) = {1\over 2\pi }\varepsilon _{jk}\partial _{j}K_{k}(\bbox{x}) .
\label{SkyrmCharg}
\end{equation}
It is a straightforward calculation \cite{Skyrmion} to show that the charge 
(\ref{SkyrmCharg}) is identical to the Pontryagin number density (\ref{PontrNumbe}).  

The topological charge is evaluated as
\begin{equation}
Q^{P} = \int \dx Q^{P}_{0}(\bbox{x}) = {1\over 2\pi i}\oint dx_{j} K_{j}(\bbox{x}) ,
\label{SkyrmChargA}
\end{equation}
where the loop integration $\oint $ is made to encircle the excitation at infinity 
($|\bbox{x}|\rightarrow \infty $).  The electron number associated with the topological soliton is
\begin{equation}
\Delta N = \int \dx \varrho  (\bbox{x}) = -\nu  Q^{P} .
\label{NumbeElectS}
\end{equation}
The topological charge is determined by the asymptotic value of the \CP field 
(\ref{GenerSkyrm}).  We find $Q^{P}=q$ for the Skyrmion (\ref{SkyrmSpin}).  The electron 
number of this soliton is $\Delta N=-\nu q$: It represents the number of electrons 
removed by the Skyrmion excitation.

The number of flipped spins is given by
\begin{equation}
\Delta N_{s}= - \int \dx \bigl\{2S_{z}(\bbox{x})-\rho _{0}\bigr\} +\Delta N = \int \dx \rho (\bbox{x})\bigl\{1-s_{z}(\bbox{x})\bigr\} ,
\label{NumbeFlipp}
\end{equation}
where $\Delta N$ is given by (\ref{NumbeElectS}).  We have subtracted it since it 
represents the number of electrons removed and not of flipped spins.  The 
necessity of the subtraction is clear in the vortex limit $\kappa \rightarrow 0$, where it 
follows that $\Delta N_{s}=0$ thanks to this term, as should be the case since no spin 
flip occurs in the vortex excitation.  

The number of flipped spins (\ref{NumbeFlipp}) would diverge logarithmically 
for the Skyrmion (\ref{SkyrmSpin}) with $q=1$.  This is an illusion since the the 
Zeeman term breaks the spin SU(2) symmetry explicitly and introduces a 
coherent length $\xi _{L}$ into the SU(2) component.  The Skyrmion configuration 
(\ref{SkyrmSpin}) is valid only within the coherent domain because the coherent 
behavior of the spin texture is lost outside it.  By cutting the upper limit 
of the integration at $r\simeq \xi _{L}$ in (\ref{NumbeFlipp}), we obtain
\begin{equation}
\Delta N_{s} = \kappa ^{2} \ln\Bigl({\xi _{L}^{2}\over \kappa ^{2}\ell _{B}^{2}}+1\Bigr)-1 ,
\label{NumbeSpinFlip}
\end{equation}
with the coherent length $\xi _{L}$ given by (\ref{CoherLengt}).

The density modulation around the Skyrmion is governed by the soliton 
equation (\ref{SolitEqSkyrm}).  This equation has formally the same expression as 
the soliton equation (\ref{SolitEqVorte}) for the vortex excitation.  We may obtain 
an approximate solution in the two limits, the large Skyrmion limit ($\kappa \gg 1$) 
and the small Skyrmion limit ($\kappa \ll 1$).  First, in the large limit we may solve 
(\ref{SolitEqSkyrm}) iteratively as
\begin{equation}
\varrho  (\bbox{x}) = -\nu Q^{P}_{0}(\bbox{x}) - {\nu ^{2}\over 8\pi \rho _{0}}\nabla ^{2}Q^{P}_{0}(\bbox{x}) + \cdots  .
\label{SolitEqSkyrmA}
\end{equation}
We may approximate it as
\begin{equation}
\varrho  (\bbox{x}) \simeq  -\nu Q^{P}_{0}(\bbox{x})={\nu \over \pi } {(\ell _{B}\kappa )^{2}\over [r^{2}+(\ell _{B}\kappa )^{2}]^{2}},\quad \quad \text{for}\quad  \kappa \gg 1 ,
\label{SolitEqSkyrmB}
\end{equation}
for the Skyrmion with $q=1$, where we have used (\ref{SkyrmPontr}).  It agrees with 
the formula due to Sondhi et al. \cite{SkyrmQH}.  We emphasize that this formula 
is valid only in the large Skyrmion limit.  On the other hand, the topological 
charge $Q^{P}_{0}(\bbox{x})$ is localized in the small limit, $Q_{0}^{P}(\bbox{x})\rightarrow q\delta (\bbox{x})$ as $\kappa \rightarrow 0$ in 
(\ref{SkyrmPontr}).  Hence, the solution is given by the vortex configuration, 
\begin{equation}
\varrho  (\bbox{x}) \simeq  -\rho _{0}\Bigl(1+{\sqrt {2}r\over \ell _{B}}-{r^{2}\over 3\ell _{B}^{2}}\Bigr)e^{-\sqrt {2}r/\ell _{B}} ,\quad \quad \text{for}\quad  \kappa \ll 1 ,
\label{SolitEqSkyrmC}
\end{equation}
which has been derived in (\ref{BetteAppro}).

We evaluate the activation energy of a Skyrmion.  In the semiclassical 
approximation it consists of the electrostatic term and the Zeeman term in 
(\ref{HamilSpinB}),
\begin{equation}
E_{\text{Skyrmion}} = {e^{2}\over 2\varepsilon }\int \dx\dy {\varrho  (\bbox{x})\varrho  (\bbox{y})\over |\bbox{x}-\bbox{y}|} + {1\over 2}g^{*}\mu _{B}B \Delta N_{s} ,
\label{SkyrmEnergSpin}
\end{equation}
with (\ref{NumbeFlipp}).  This can be calculated by using (\ref{SolitEqSkyrmB}) and 
(\ref{NumbeSpinFlip}) for a large Skyrmion, 
\begin{equation}
E_{\text{Skyrmion}} = {e^{2}\over \varepsilon \ell _{B}}\Bigl[{\beta \nu ^{2}\over \kappa }+ {\widetilde{g}\over 2}\kappa ^{2} \ln\Bigl({\xi _{L}^{2}\over \kappa ^{2}\ell _{B}^{2}}+1\Bigr)-{\widetilde{g}\over 2}\Bigr],
\label{SkyrmEnergSS}
\end{equation}
with $\beta =3\pi ^{2}/64$.  For a sufficiently small Skyrmion where $\Delta N_{s}=0$, the Coulomb 
energy is calculated with the vortex configuration and is given by 
(\ref{CouloEnerg}).

The Coulomb energy increases for a smaller Skyrmion while the Zeeman 
energy increases for a larger Skyrmion.  The optimized scale $\kappa $ is obtained 
so as to minimize the total energy (\ref{SkyrmEnergSS}),
\begin{equation}
\kappa  \simeq  \beta ^{1/3}\bigl(\widetilde{g}\ln(\widetilde{g}^{-1})\bigr)^{-1/3} ,
\label{OptimScale}
\end{equation}
for $\widetilde{g}\ll 1$.  It yields $\kappa \simeq 1.8$ for $\widetilde{g}=0.02$.  The Skyrmion scale is of the 
order of $1.8\ell _{B}$ at 10 Tesla.  The number of flipped spins is calculated by 
(\ref{NumbeSpinFlip}), as gives $\Delta N_{s}\simeq 4.7$ for $\widetilde{g}=0.02$.  The Skyrmion activation 
energy is calculated by (\ref{SkyrmEnergSpin}), which we illustrate in 
Fig.\ref{SkyrEneUPS}.  
\FigOut{\FigSkyrEneU}

A Skyrmion is a quasihole and an anti-Skyrmion is a quasielectron in 
the QH ferromagnet.  We can make a theory of anti-Skyrmions from its 
microscopic wave function just as we have constructed a theory of 
antivortices.  Though a detailed theory is yet to come, it is clear that the 
large scale structure is the standard one, which is obtained just by reversing 
the sign of $s^{y}$ in (\ref{SkyrmSpin}),
\begin{equation}
s^{x} = \sqrt {1-(s^{z})^{2}}\cos(q\theta ),\quad
s^{y} =\sqrt {1-(s^{z})^{2}}\sin(q\theta ),\quad
s^{z} = {r^{2q}-(\ell _{B}\kappa )^{2q}\over r^{2q}+(\ell _{B}\kappa )^{2q}}.
\label{SkyrmASpin}
\end{equation}
The Skyrmion and anti-Skyrmion excitations are illustrated in 
Fig.\ref{SkyrExciPS}(b).  We expect that the activation energy is nearly the 
same as that of the Skyrmion because they have the same electric charge except 
their signs.  The activation energy $\Delta $ of a quasihole-quasielectron pair is 
observed experimentally by measuring the longitudinal resistivity 
(\ref{LongiResis}).  In Fig.\ref{SkyrEneUPS} we compare our theoretical result with 
the experimental data obtained by Schmeller et al. \cite{SkyrmExper}.
\FigOut{\FigSkyrEneT}

Our estimations are about two times bigger than the observed data 
\cite{SkyrmExper}.  This would be due to an oversimplification by approximating 
the quantum well by an infinitely thin layer.  It is correct that the motion 
of electrons into the $z$ axis is frozen completely at sufficiently low 
temperature.  However, it is not justified to assume that electric charges are 
localized within an infinitely thin layer.  The width of the layer is of the 
same order of the vortex-core size $\sim \ell _{B}$.  Its effect reduces the Coulomb 
energy by spreading the charge into a wider domain.  The best fit is obtained 
by replacing $\alpha _{C}$ with $0.46\alpha _{C}$ in (\ref{CouloEnerg}), as is shown in 
Fig.\ref{SkyrEneTPS}.  The number of flipped spins would be $\Delta N_{s}\simeq 3.2$ at 
$\widetilde{g}=0.02$, as is also a reasonable estimation compared with the data 
\cite{SkyrmExper}.
\section{Discussion}

We have presented an improved \CB theory of QH states, where the field 
operator describes solely the physical degree of freedom representing the 
deviation from the ground state.  We have successfully analyzed excited states 
as well as the ground state.  In this scheme the semiclassical properties of 
topological solitons, vortices and Skyrmions, are closely related to their 
microscopic wave functions.

The improved \CB theory is based on the bosonic field operator $\varphi (\bbox{x})$ 
defined by (\ref{DressField}).  Such an operator was first considered by Read 
\cite{ReadA} to construct a Landau-Ginzburg model of QH effects.  However, his 
effective theory is quite complicated and so different from our microscopic 
theory.  This operator was revived by Rajaraman et al. \cite{RajaramanCB}.  
Although the operator itself is identical, our conclusions are significantly 
different from theirs.  According to their conclusions, their formalism is 
not a unitary theory, it is useful without the LLL projection and only the 
ground state is successfully analyzed.  On the contrary, our theory is unitary 
together with an integration measure, and the LLL projection has played the 
key role.  First, it transforms the Abelian translational group into the 
magnetic translational group, as is the origin of the Coulomb exchange energy 
and leads to the NL$\sigma $ model describing the Goldstone mode in the QH 
ferromagnet.  Second, we have introduced vortices and Skyrmions merely as 
excitations allowed in the \LLL.  They induce electron density modulations 
according to the soliton equation which is the semiclassical LLL condition.

Skyrmions are generic SU(2) excitations in the \LLL.  The classical 
configuration is determined from its microscopic wave function.  This is to be 
contrasted with the standard approach \cite{SkyrmQH}, where it is introduced as a 
classical solution in the NL$\sigma $ model (\ref{EnergChangSPN}).  Thus, the Skyrmion 
semiclassical energy is
\begin{equation}
E^{\text{ours}}_{\text{Skyrmion}} = E_{C} + E_{Z} 
\label{OurFormu}
\end{equation}
in our theory, where $E_{C}$ and $E_{Z}$ is the semiclassical Coulomb and Zeeman 
energies in (\ref{SkyrmEnergSpin}), while it is
\begin{equation}
E^{\text{theirs}}_{\text{Skyrmion}} = {1\over 2}\rho _{s}\sum _{a}\int \dx [\partial _{k}s^{a}(\bbox{x})]^{2} + E_{C} + E_{Z} 
\label{TheirFormu}
\end{equation}
in the standard literature \cite{SkyrmQH,MG}.  We make comments on this point.  As 
we have stated, there are two complementary methods to estimate the energy 
$\langle \F|H|\F\rangle $ of the spin texture.  The algebraic method is appropriate for the 
analysis of small perturbative fluctuations around the ground state, where the 
NL$\sigma $-model term has been derived \cite{EzaIQC,MG} in the lowest order 
approximation.  On the other hand, the semiclassical approximation is a 
powerful method for the analysis of nonperturbative excitations, where there 
is no reason to include the NL$\sigma $-model term.  Indeed, its absence is required 
from the consistency condition that the Skyrmion wave function is reduced to 
the vortex wave function in the limit $\kappa \rightarrow 0$.  We have checked that, as $\kappa \rightarrow 0$, 
the Skyrmion excitation energy is reduced precisely to the vortex excitation 
energy based on our formula (\ref{OurFormu}).  We also point out that, accordind to 
their formula (\ref{TheirFormu}), due to the NL$\sigma $-model term the Skyrmion 
excitation energy would become larger than the vortex excitation energy for 
almost all values of $\widetilde{g}$, as is seen in Fig.\ref{SkyrEneTPS}.  This means that 
Skyrmion would not be relevant excitations, as contradicts experimental 
observations of Skyrmions.

I would like to thank A. Sawada, I. Takagi and K. Tsuruse for various 
discussions on the subject.  A partial support is acknowledged from a Grant-
in-Aid for the Scientific Research from the Ministry of Education, Science, 
Sports and Culture.

\end{document}